\documentclass[journal=jpccck,manuscript=article]{achemso}
\setkeys{acs}{articletitle = true}
\usepackage[usenames,dvipsnames]{xcolor} 
\usepackage{changes} 
\usepackage{hyperref} 
\usepackage{url} 
\usepackage{float} 
\usepackage[font=scriptsize]{caption} 
\usepackage{tabularx} 
\usepackage[utf8]{inputenc}
\usepackage[T1]{fontenc}
\usepackage[version=4]{mhchem}
\usepackage{mwe} 
\usepackage{subfig} 
\usepackage{silence}
\usepackage{array}   
\usepackage{graphicx}
\usepackage{multirow} 
\usepackage{booktabs}

\title{Combining DFT with ML to study size specific interactions between metal clusters and adsorbates}

\author{\rm Shweta Mehta}
\affiliation{ Physical and Materials Chemistry Division, 
              CSIR-National Chemical Laboratory,
	      Dr. Homi Bhabha Road, Pashan, Pune-411008, India.}
\altaffiliation{Academy of Scientific and Innovative Research (AcSIR) }
\author{\rm Sheena Agarwal}
\affiliation{ Physical and Materials Chemistry Division, 
              CSIR-National Chemical Laboratory,
	      Dr. Homi Bhabha Road, Pashan, Pune-411008, India.}
\altaffiliation{Academy of Scientific and Innovative Research (AcSIR) }
\author{\rm Kavita Joshi}
\affiliation{ Physical and Materials Chemistry Division, 
              CSIR-National Chemical Laboratory,
	      Dr. Homi Bhabha Road, Pashan, Pune-411008, India.}
\altaffiliation{Academy of Scientific and Innovative Research (AcSIR) }
\email{k.joshi@ncl.res.in; kavita.p.joshi@gmail.com} 
\date{\today}


\begin{document}
\begin{abstract}
To date, density functional theory (DFT) is one of the most accurate and yet practical theory to gain
insight about materials properties. Although successful, the computational cost is the main hurdle 
even today. A way out is combining DFT with machine learning (ML) to reduce 
the computational cost without compromising accuracy.  
However, the success of this approach hinges on the correctness of the descriptors. In the present work,
we demonstrate that, based on {\it only} interatomic distances as descriptors, our ML model predicts interaction energy between 
an adsorbate and Al cluster with absolute mean error (AME) $\sim$ 0.05  eV (or less) and reproduces
the PES experienced by an incoming atom.  Our extensive DFT calculations 
reveal that atoms experiencing identical environment within a cluster have identical interaction energy patterns. 
Further, we demonstrate that our model is not specific to Al 
clusters, and could be applied to clusters of different elements as well. Its application to compute PES experienced 
by various test atoms and molecules in the vicinity of different clusters proves the transferability of the model 
not just to clusters of different elements but also to various molecules. The descriptors chosen are invariant 
to rotation, translation, and permutation yet very simple to compute is one of the most crucial points of the present work.
\end{abstract}
\maketitle 
\section{Introduction}
Clusters, being the system of few atoms, have very different properties than their corresponding atomic
and macroscopic analogue\cite{jortner1992, AuLandman, PNAS2006, castleman2009, alonso2012structure, berry2013bridging}. 
Due to the unique arrangement of atoms, the properties of clusters vary substantially with size. 
This size sensitivity is the key feature of atomic clusters and is reflected in all their properties 
like, melting\cite{schmidt1998, jarrold2004, kavita2006, berry2013}, their growth pattern\cite{jarrold1999}, reactivity towards 
various molecules\cite{cox1988, 
argo2002, fu2003, campbell2004, chen2004, 
lemire2004, wei2004, vajdaPt}, etc. The reactivity of clusters has been investigated by both experimental and
theoretical means\cite{heiz1998, wallace2000, jarroldjacs, roach2009, reber2010, kulkarni2011, yin2012gas, review2016}.
Roach {\it et al.} studied the reaction of anionic Al$_{7-73}$  clusters with H$_{2}$O molecule\cite{roach2009, reber2010}. 
They demonstrated that Al$_{12}$$^{-}$ cluster adsorbs more than one H$_{2}$O molecule, while Al$_{13}$$^{-}$
does not react with H$_{2}$O. 
The electronic closed shell structure (40 electrons) was considered to be the reason for non-reactivity of Al$_{13}$$^{-}$. 
But, Al$_{23}$$^{-}$ which is also a closed shell (70 electrons) does react while Al$_{20}$$^{-}$ which is not a closed 
shell (61 electrons) does not react with H$_{2}$O. Hence this variation in reactivity has geometric rather than electronic origin .
By adding an atom in Al$_{12}$$^{-}$, the reactivity completely diminishes due to 
the absence of adjacent Lewis acid-Lewis base active sites in Al$_{13}$$^{-}$ cluster.
In smaller size regime where every atom counts, addition or removal of just an atom dramatically 
changes its properties. Thus, it becomes very difficult to bring out general trends in this size range.  
It has been also demonstrated by a few groups that not just size, site also affects the reactivity of clusters
\cite{pal2013, castleman2013}. And hence, to model the interaction of clusters with an incoming adsorbate, 
all possible adsorption sites for all the clusters must be scanned, which in turn leads to a prohibitively large number of DFT calculations. 
To overcome this problem, we have employed 
data driven algorithms of machine learning to predict the site specific interaction energies for various aluminum clusters.
Machine learning is preferred over conventional curve fitting because it brings out the underlying complex relations buried 
in the data set which are useful in prediction for newer and unseen data points as we will demonstrate
in this work.

Use of ML methods in combination with DFT is continuously increasing in the field of
materials\cite{MLinmatsci, butler2018machine}. It is being applied for accelerated materials discovery
\cite{ternaryoxcomp, acceleratingmatdiscov, combinatorialscreen}, to understand the underlying 
electronic structure\cite{dbandcen}, to obtain chemical information\cite{rxnnetwork, bose2018machine}, 
to predict the potential energy functions \cite{bukkapatnam2006parametrization, JPCA2013watercluster, RCS2015water, QuantChem2015rev, JCP2016permutation, 
dragoni2018achieving, jeong2018toward, zhang2018potential} and so on.
Success of any ML model hinges on choosing the right set of descriptors. Descriptors should be such that they
can bring out accurate and hidden trends from a data set, and yet be as simple as possible. Thus designing 
features/descriptors that completely describe the system are very crucial. 
Lot of efforts are being put in developing fingerprints that systematically relate the
structural features of samples to their functional properties in quantitative terms\cite{le2012quantitative, schutt2014represent, von2015fourier, seko2017representation}. These set of features 
then find varied applications like finding similarity between two structures
\cite{rupp2012fast, faber2017prediction, bartok2017machine}, 
finding the structure-activity relation for various systems
\cite{davran2010structure, jager2018machine, musil2018machine}, 
screening the chemical space to discover novel materials of desired properties\cite{hansen2015, xie2018crystal} or 
even predict properties for a given material
\cite{pilania2013accelerating, montavon2013machine,
 rupp2015machine, chandrasekaran2019solving}.  
In a study by Hansen {\it et al.}, they outlined a number of established machine learning techniques and investigated 
the influence of the molecular representation on ML methods performance. 
The best methods achieve prediction errors of 3 kcal/mol (0.13 eV) for the atomization energies of a wide 
variety of molecules\cite{hansen2013}.

Few groups have recently used ML for {\it in silico} design of catalysts and proved the validity of their 
model against the first principles methods\cite{Co2chemisorp, bimetcat, featureengg}. 
Wang {\it et al.} recently demonstrated the use of artificial neural networks combined with kinetic analysis 
for rapid screening of bimetallic catalysts. Through a Machine Learning model, they could capture the underlying 
complex and non-linear interaction between adsorbate and metal, with reported RMSE $\sim$ 0.2 eV\cite{bimetcat}.
Another group, Ma {\it et al.} adopted the use of ML to capture interactions 
of adsorbate on multimetallics for catalyst screening of CO$_{2}$ electroreduction with an RMSE of $\sim$ 0.1 eV\cite{Co2chemisorp}. 
Other studies integrating {\it ab initio} calculations and ML, for transition metal catalysts screening have 
reported errors (RMSE) as low as $\sim$ 0.12 eV\cite{featureengg}. 

Owing to the excellent catalytic properties, nanoparticles and atomic clusters have always been objects of interest\cite{informatics, li2017application, predcatnano, CoonPt, kitchin2018machine}. 
Recently, there is an increasing trend in resorting to ML for discovering correlations between geometric structure 
and catalytic activities\cite{li2017application, kitchin2018machine}  of metal surfaces as well as nanoparticles\cite{informatics}. 
Recently an ML scheme was proposed to understand catalytic activities based on local atomic 
configurations and applied to study direct NO decomposition on RhAu alloy nanoparticles\cite{predcatnano}. 
A local structural similarity kernel known as a smooth overlap of atomic positions (SOAP)
was used to find similarities between two geometries based on structural descriptors. 
Gasper {\it et al.} used the gradient-boosting algorithm, for prediction of CO adsorption energies on Pt clusters\cite{CoonPt}.
They built predictive models of site-specific adsorbate binding on 
realistic, low-symmetry nanostructures, with AME $\sim$ 0.1 eV (with respect to DFT). 
Descriptors used during the training of the ML model in this study comprised of
d-band center energy, s and p band center energies, Bader charges, generalized coordination number, etc. 

In the present work, we use both DFT and ML techniques, to predict the interaction energy of Al clusters with H atom as an adsorbate. 
This interaction is studied as a function of increasing size. DFT investigations bring out the one to one correlation
between neighbor distance distribution of atoms in a cluster and their corresponding interaction energy. This strong 
correlation provides the rationale of choosing distances between adsorbate and the surface atoms of cluster as descriptors
to train the ML model. And indeed our model based on `only' distances as descriptors could predict the interaction
energy with errors as low as 0.05 eV. Further, the transferability of our ML model is demonstrated by its application to different 
homogeneous (Na$_{10}$) as well as bimetallic clusters (Al$_{6}$Ga$_{6}$). To validate our model the adsorbate is also 
replaced by other atoms (N), and molecules (N$_{2}$, O$_{2}$, and CO).
\section{Computational Details}
To determine the overall reactivity of a cluster, site specific interaction energy needs to be evaluated. 
However, owing to the lack of long range order
and highly altered short range order, different atoms within a cluster interact
differently with an incoming adsorbate. 
To quantify this site specific variation, we have computed
interaction energy of various atoms like H, N and molecules like N$_{2}$, O$_{2}$, and CO with Al clusters.
\begin{figure}[h]
  \centering
   \includegraphics[scale=0.39]{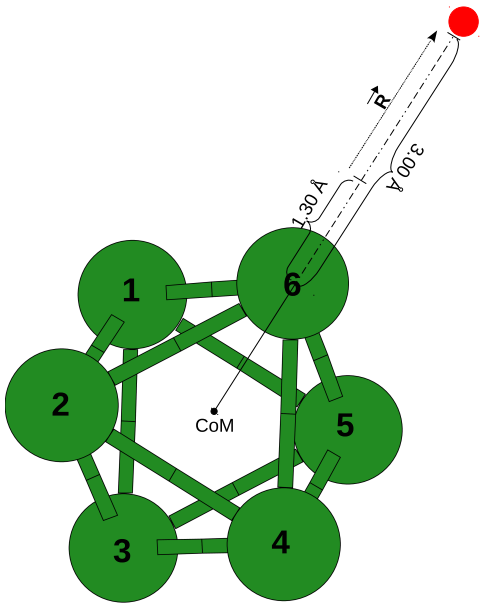}
   \caption{Al$_{6}$ cluster is shown with center of mass marked at the center. The adsorbate is placed along the radial 
vector pointing outwards from the center of mass to surface atom. Distance between adsorbate and surface atom is varied from 1.30 \AA{} to 3.00 \AA.}
  \label{fig1}
\end{figure}
Also, the interaction of H atom with cluster of another element like Na$_{10}$ and bimetallic cluster like Al$_{6}$Ga$_{6}$
was computed. All these resulted into about 35,000 single point calculations. 
The adsorbates were placed at on-top position of all the surface sites
(i.e. surface atoms) for all the clusters with size ranging from 5 to 80. The GS geometries for all the clusters 
were taken from previously reported work\cite{neal2007melting, aguado2008, aguado2009, anju2014}.
As shown in Fig.\ \ref{fig1}, the adsorbate was kept along the outward radial vector 
from center of mass of the cluster to surface atom. The distance of adsorbate was varied between 1.30 \AA{} to 3.00 \AA{}
from the surface site. 
All the calculations were carried out within the Kohn-Sham formulation
of DFT. Projector Augmented Wave pseudopotential
\cite{paw1,paw2} was used, with Perdew--Burke--Ehrzenhof
(PBE) \cite{pbe1} approximation for the exchange-correlation and generalized gradient
\cite{pbe2} approximation, as implemented in
plane wave pseudo potential based code, \textit{VASP}\cite{vasp1,vasp2,vasp3}.
Cubic simulation cell, with the image in each cell separated by at least 15 \AA~ of vacuum, was used.
Energy convergence criteria of 10$^{-5}$ eV was used for SCF calculations.

Data collected from DFT calculations was then used to train a ML model.
We used the Gradient Boosting Regression (GBR) algorithm as implemented in the
scikit-learn python package\cite{scikit}.
 GBR is a regression technique that uses 
decision tree based classifiers as weak learners.  We used the mean squared error function 
as our loss function (i.e. the objective function to be optimized).  
The GBR was selected after comparing it against four other regression algorithms viz.
Linear Regression, Ridge Regression, LASSO and Stochastic Gradient Descent (SGD).
An exhaustive grid search was carried out to find the best parameter values of an estimator.
 5-fold cross validation was performed to test accuracy of the model.
AME  was used as the scoring parameter during cross validation. 
Multiple checks like plotting the validation curve and learning curves were used to 
ensure that the model did not overfit the data.
\section{Results and Discussion}
In clusters, due to the finiteness of the system, every atom does not experience identical environment unlike 
atoms in bulk. To quantify this variation, nearest neighbor distribution of every atom in all the clusters was 
studied. In this distribution,
distance (d$_{i,j}$) between every pair of atom i and j in a cluster was calculated. 
\begin{figure}
  \centering
   \includegraphics[scale=0.76]{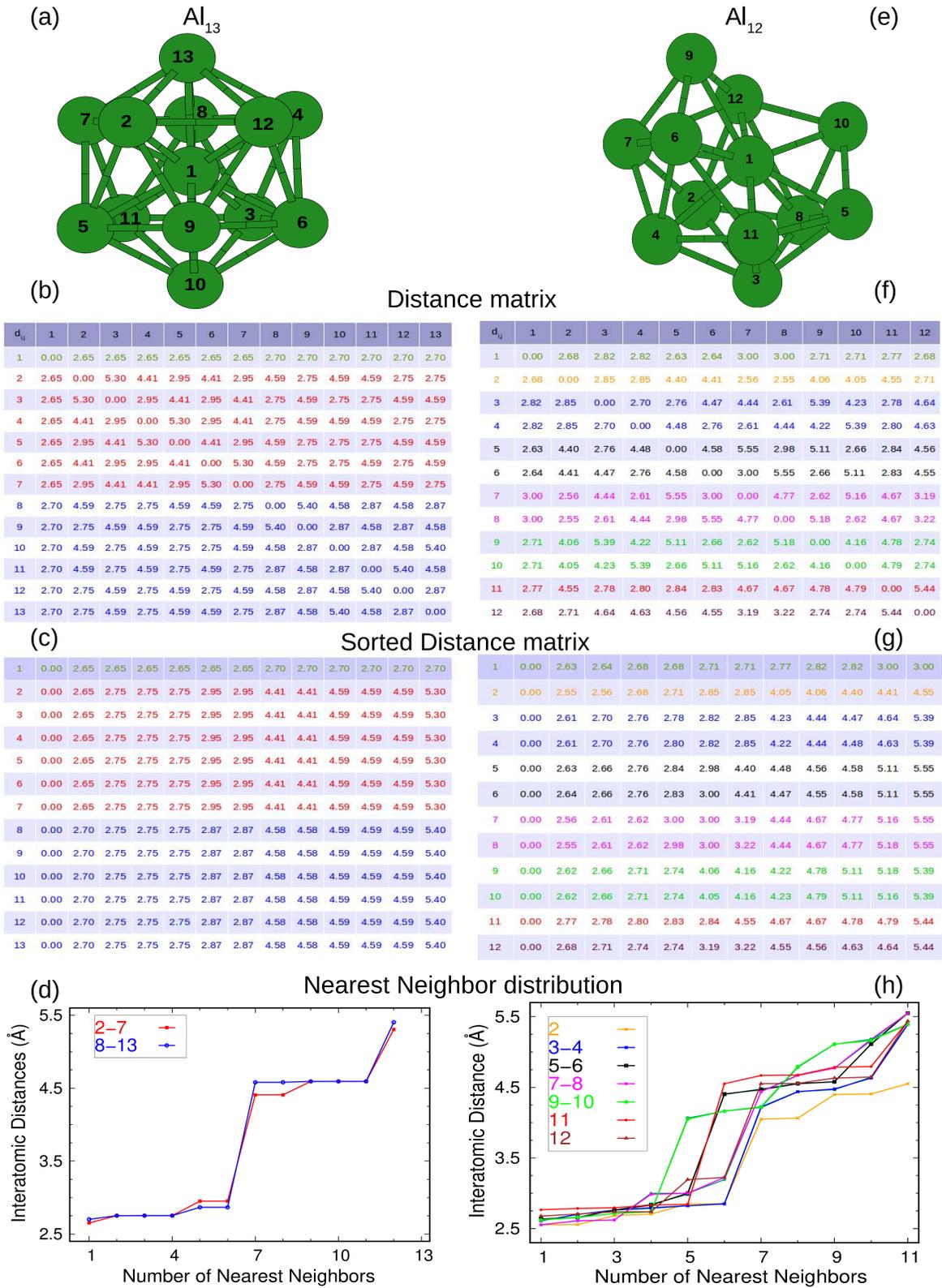}
  \caption{(a) Geometry of Al$_{13}$ cluster, (b) Distance matrix(DM) of Al$_{13}$ 
where atoms experiencing identical environment are shown in same color, (c) Sorted Distances in each row of DM, (d) Nearest Neighbor (NN) distribution for Al$_{13}$ showing two distinct lines.  
(e) Geometry of Al$_{12}$ cluster, (f) DM of Al$_{12}$ where atoms 
experiencing identical environment are shown in same color, (g) Sorted Distances in each row of DM,
(h) NN distribution for Al$_{12}$ showing seven distinct lines. NN distribution for the core atoms is not shown.}
  \label{fig2} 
\end{figure} 
Fig.\ \ref{fig2}-(a) shows the geometry of Al$_{13}$ cluster.
The corresponding $13 \times 13$ distance matrix for Al$_{13}$ is shown in Fig.\ \ref {fig2}-(b). To identify the atoms
that experienced identical environment (in terms of neighbor distances), distances were arranged in ascending order 
in the sorted distance matrix as shown in Fig.\ \ref {fig2}-(c). A careful look at this matrix revealed that 
there were only two unique rows. Implying
that all the twelve surface atoms in Al$_{13}$ were grouped in two classes as is also seen in Fig.\ \ref{fig2}-(d).
Similarly, for Al$_{12}$ cluster, eleven surface atoms were grouped into seven classes as shown in Fig.\ \ref{fig2}-(e-h).
It should be noted that upon addition of just one atom in Al$_{12}$ cluster, seven different classes merge and form only two classes
in the case of Al$_{13}$, i.e. an asymmetric cluster gets transformed into a highly symmetric one.
As evident from the Fig.\ \ref{fig2}-(d) and Fig.\ \ref{fig2}-(h), the interatomic distances can be used to indicate how (dis)ordered
the cluster is. By `ordered' cluster we mean a cluster with many identical atoms in terms of the chemical environment that they 
experience. For example, Al$_{13}$, and Al$_{36}$ are `ordered' clusters because all the surface atoms 
are grouped into 2 for Al$_{13}$ as shown in Fig.\ \ref{fig2}-(d), and 6 for Al$_{36}$ classes as shown in Fig.\ \ref{fig3}-(a). Whereas for all the disordered clusters, more than half of surface atoms 
experience unique environment like 7\ in case of Al$_{12}$ as shown in Fig.\ \ref{fig2}-(h) and 23\ in Al$_{25}$ as shown in Fig.\ \ref{fig3}-(b).
\begin{figure}[h]
  \centering
  \begin{minipage}[b]{0.45\linewidth}
   \centering
   \textbf{Ordered Cluster}\par\medskip
   \includegraphics[width=0.99\textwidth]{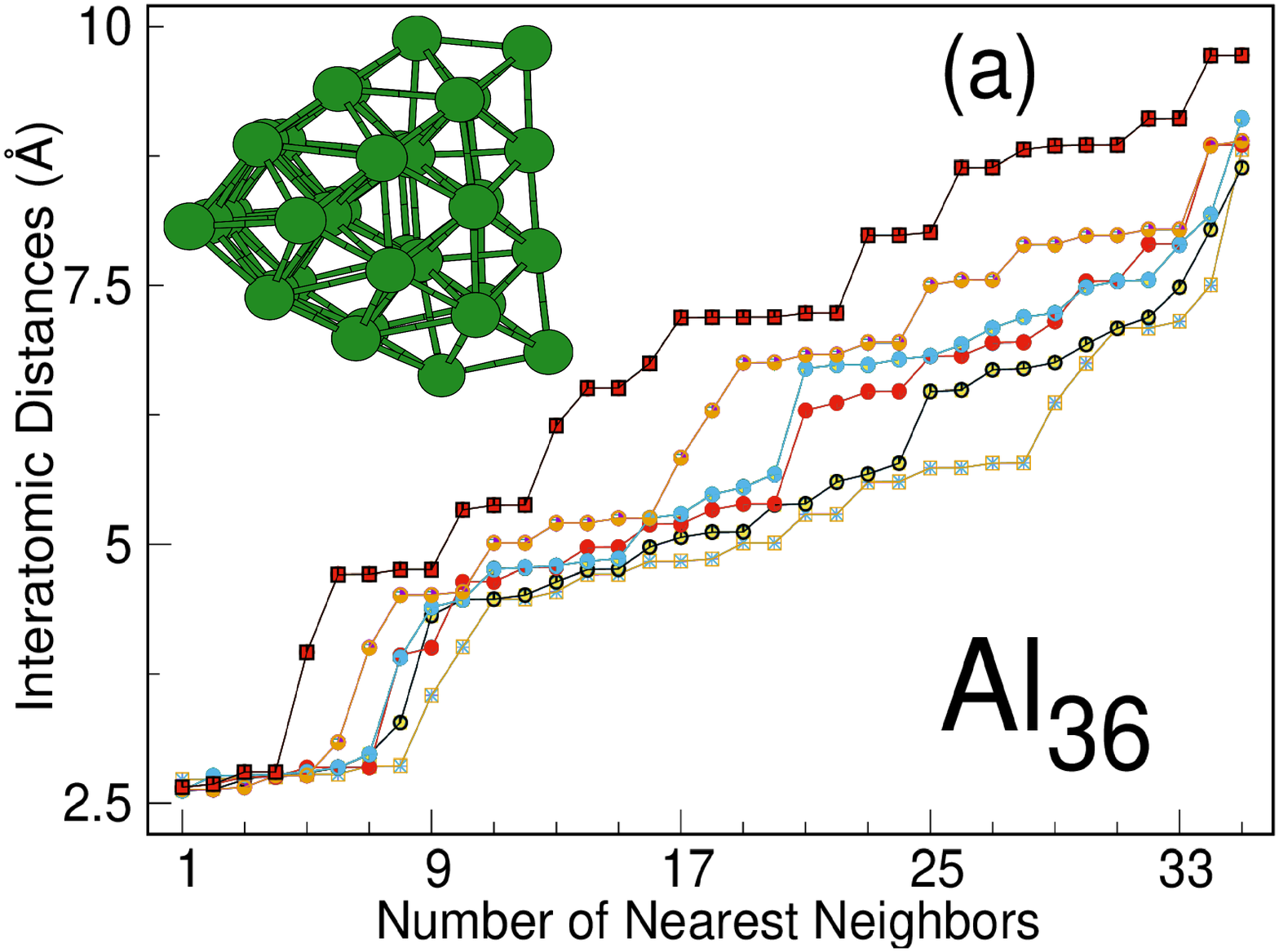}
  \end{minipage}
  \begin{minipage}[b]{0.45\linewidth}
   \centering
   \textbf{Disordered Cluster}\par\medskip
    \includegraphics[width=0.99\textwidth]{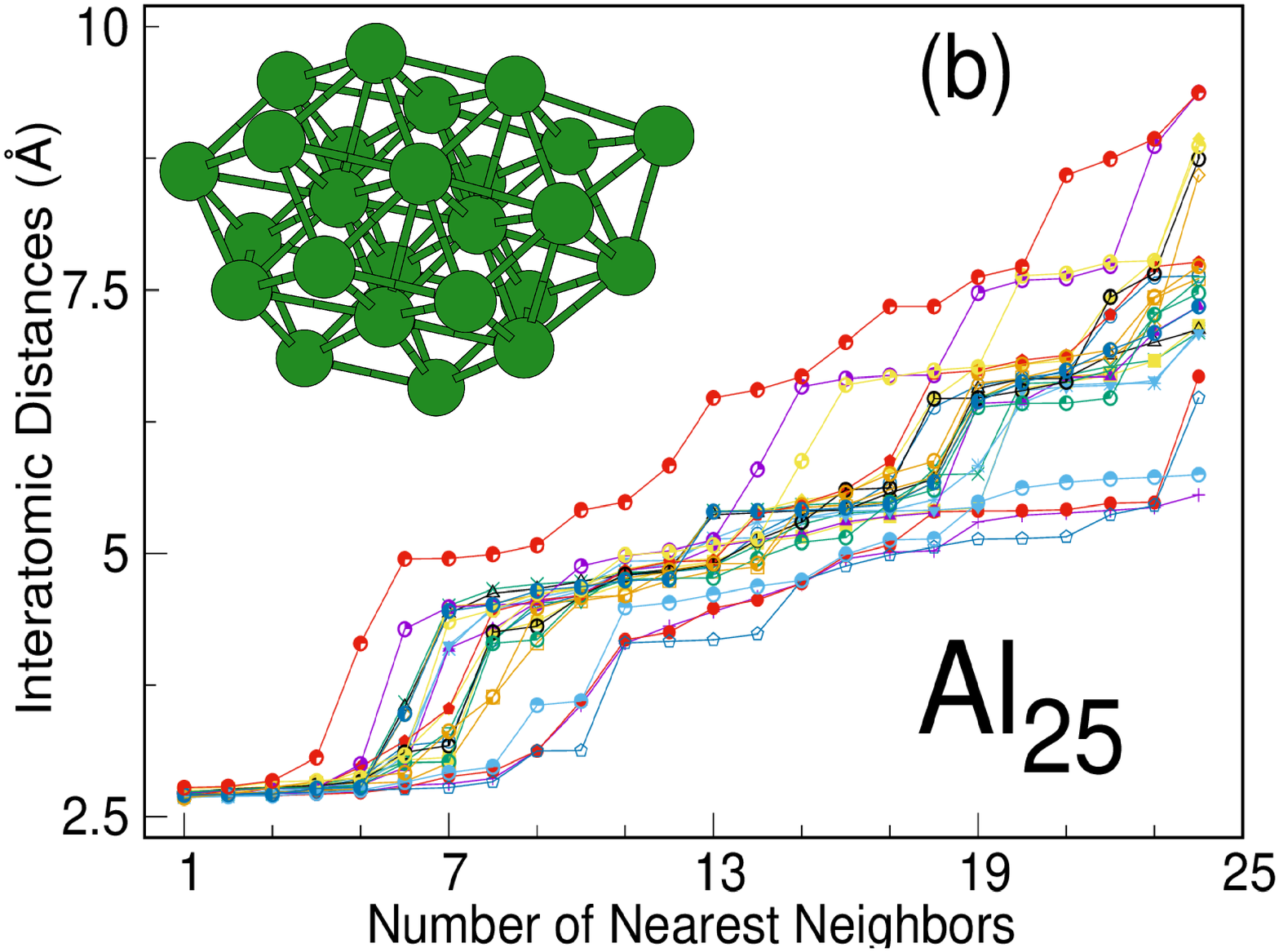}
  \end{minipage}
  \caption{Figure shows variation in the interatomic distances as function of nearest neighbors for all 
surface atoms. (a) shows the `ordered' Al$_{36}$ cluster, having 32 surface atoms which are grouped into six different classes. 
(b) shows `disordered' Al$_{25}$ cluster. In this cluster, there are 23 surface atoms, all of 
these atoms have different distance distribution. Only the distances of surface atoms are shown in NN distribution.
The inset figure shows the geometry of respective clusters.}
  \label{fig3}
\end{figure}

This variation in the arrangement of individual atoms is a characteristic of atomic clusters in this size regime. 
As noted earlier, Roach {\it et al.} demonstrated a substantial change in the reactivity of Al$_{12}$$^{-}$ compared to
Al$_{13}$$^{-}$ cluster towards H$_{2}$O molecule\cite{roach2009, reber2010}. A proof of concept for such experimental studies on reactivity 
lies in our observation based on a significant change in the symmetry of clusters with addition of just one atom 
as shown in Fig.\ \ref{fig2}-(d) and Fig.\ \ref{fig2}-(h). It brings out the fact that variation in behavior of clusters with changing size 
originates from their geometries. And hence, motivation of our work lies on understanding the interaction of
clusters with incoming adsorbates as a function of changing geometries with size (fixed geometry for a size).
In our work, cluster geometries were expressed in terms of the nearest neighbor distribution.
We demonstrate that the site specific interaction depends upon the nearest neighbor distribution of that 
specific atom (or site) within a cluster. 

\begin{figure}[h]
  \begin{minipage}[b]{0.48\linewidth}
   \includegraphics[width=\textwidth]{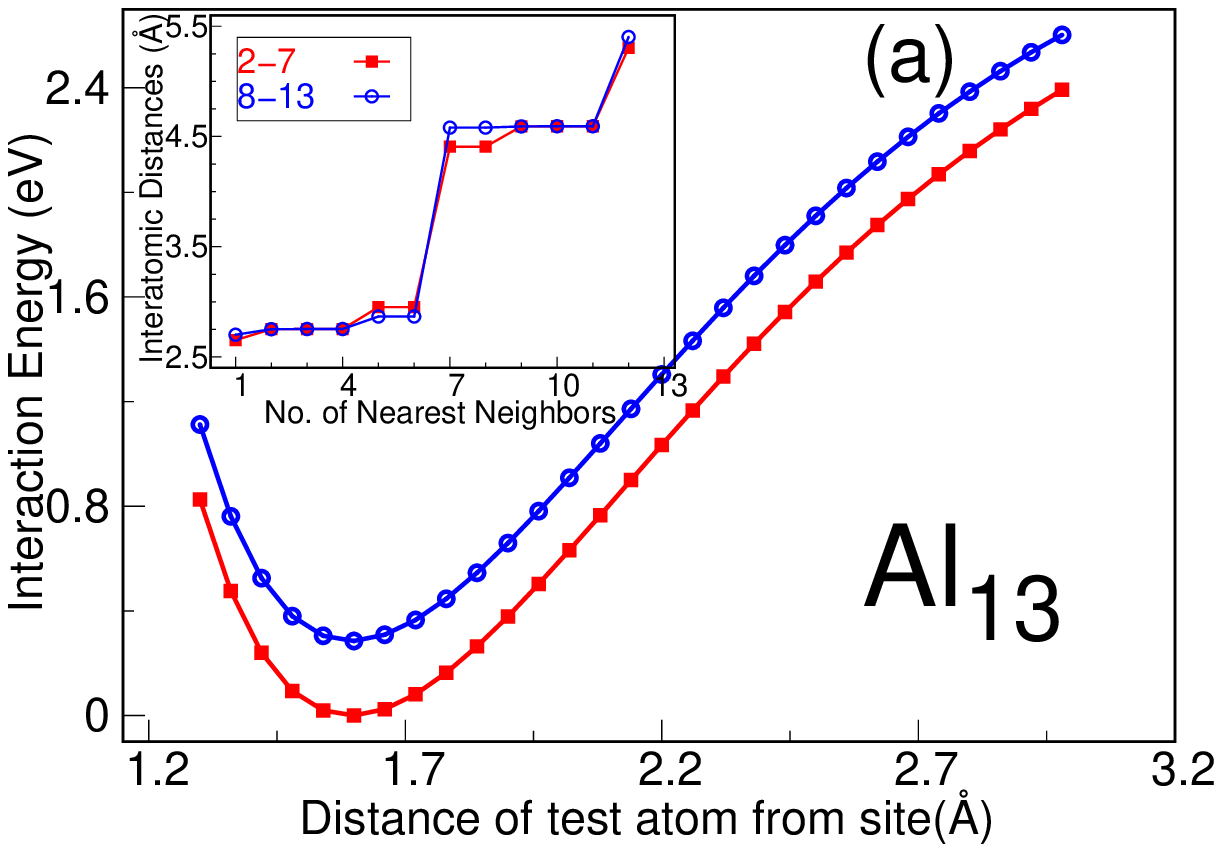}
  \end{minipage}
  \begin{minipage}[b]{0.48\linewidth}
   \centering
   \includegraphics[width=\textwidth]{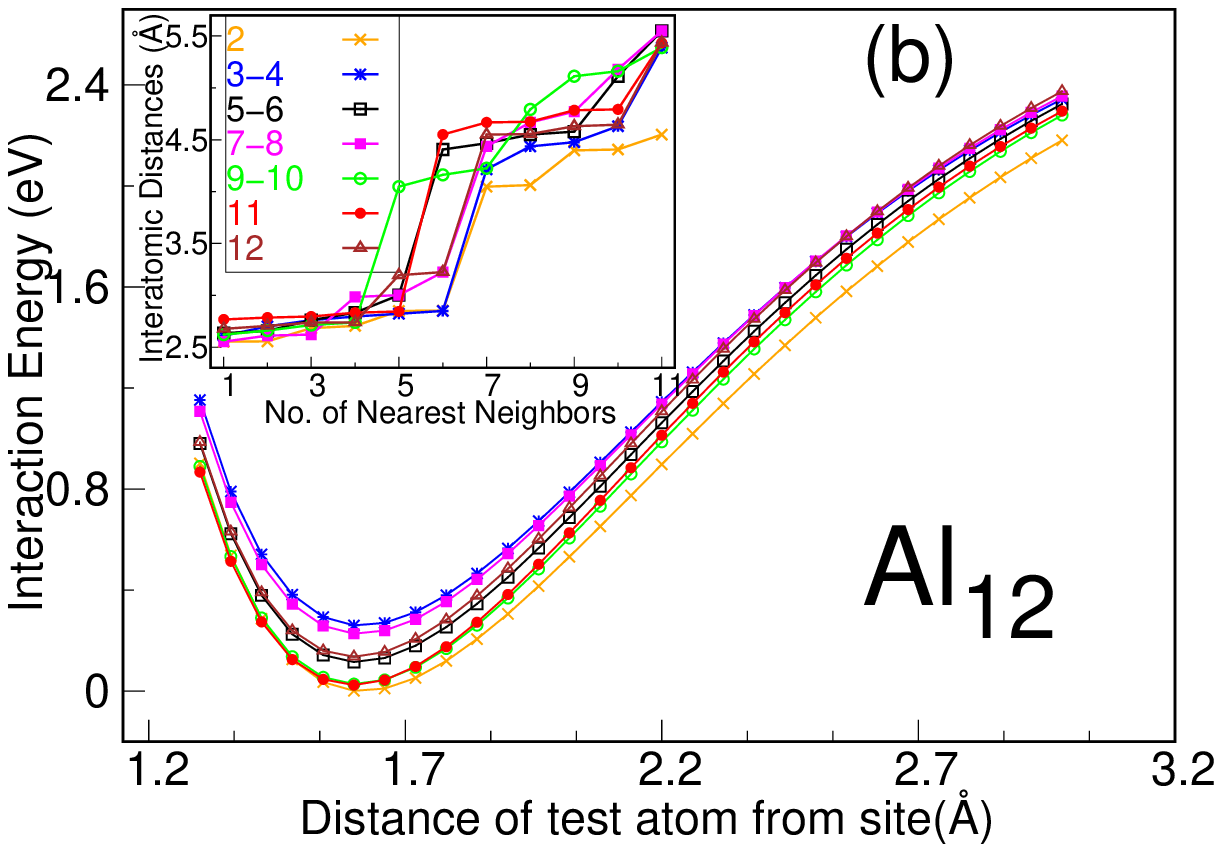}
  \end{minipage}
\vfill
  \begin{minipage}[b]{0.48\linewidth}
   \centering
   \includegraphics[width=\textwidth]{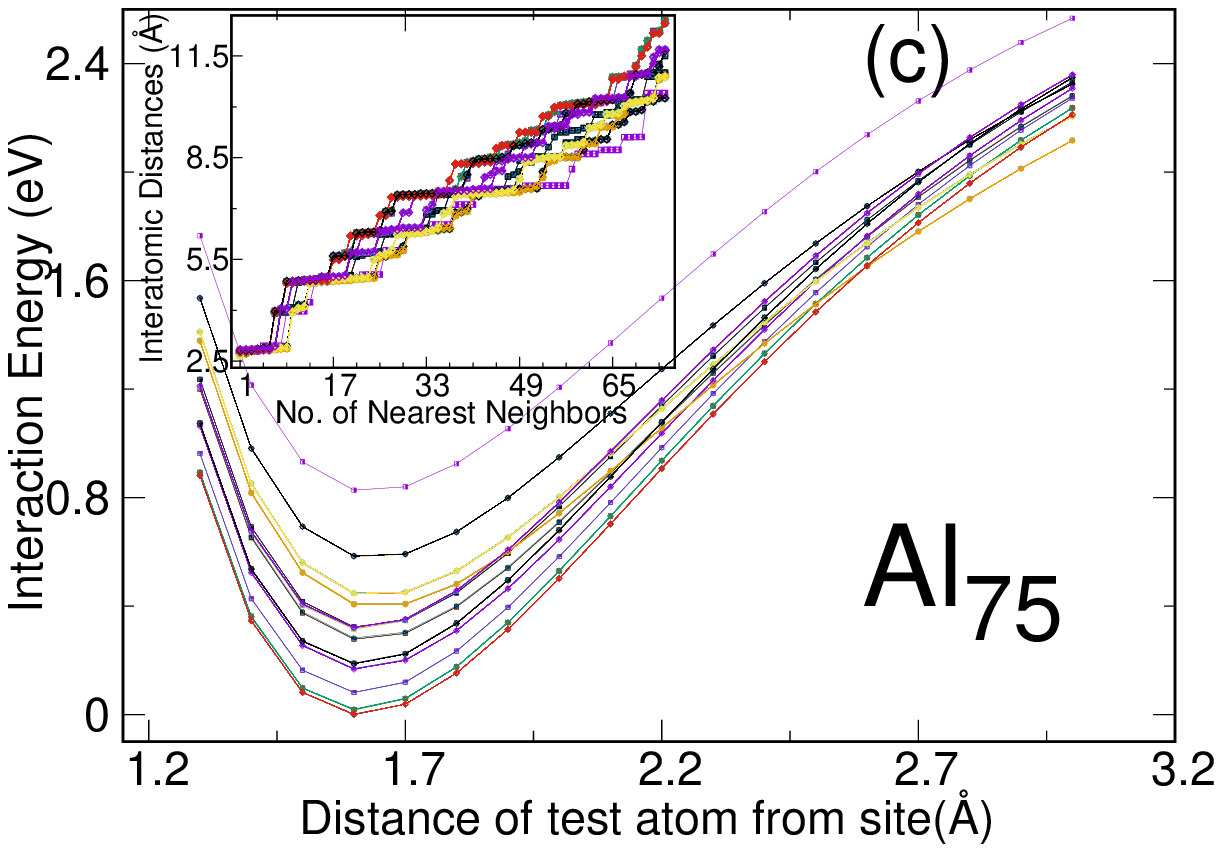}
  \end{minipage}
  \begin{minipage}[b]{0.48\linewidth}
   \centering
   \includegraphics[width=\textwidth]{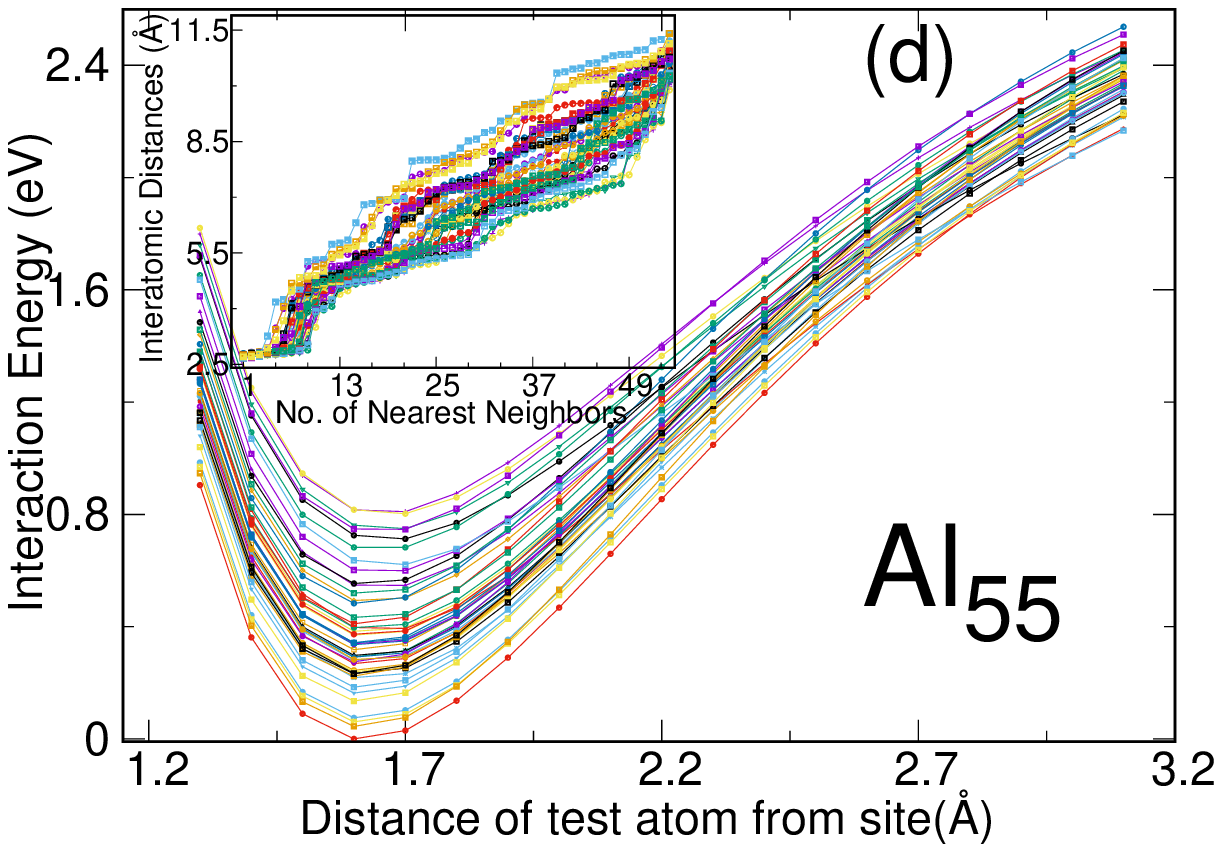}
  \end{minipage}
  \caption{The figure shows distance dependent interaction energy of all surface atoms for a few representative clusters.
Atoms with identical nearest neighbor distribution also exhibit identical interaction energy pattern towards an H atom. 
Inset figure shows variation in the interatomic distances as function of nearest neighbors for all surface atoms of these clusters.}
  \label{fig4}
\end{figure}
All atoms having identical nearest neighbor distribution 
within a cluster, interact identically with the incoming adsorbate. To elaborate this point further, in Fig.\ \ref{fig4} 
we show the interaction energies for all the atoms within a cluster,  for a few representative sizes 
along with their nearest neighbor distribution (or interatomic distances) in the inset. In the case of Al$_{13}$ 
(see inset of Fig.\ \ref{fig4}-a) as explained earlier all the surface atoms could be grouped in two classes based on their 
respective interatomic distances, indicating that an incoming adsorbate would experience only `two' different 
environments. Further, when interaction energy of these surface atoms with an H atom (as adsorbate) was 
computed, it was observed that atoms belonging to one class interact identically with the adsorbate, resulting into identical 
interaction energy as shown in Fig.\ \ref{fig4}-(a).
This is evident from the interaction energy of the adsorbate with all the surface atoms when placed at `on top' position. 
This one to one correlation between identical nearest neighbor distribution and interaction energy is also observed in larger clusters like Al$_{75}$ and Al$_{55}$ as shown in Fig.\ \ref{fig4}-(c) and \ \ref{fig4}-(d) respectively.
\begin{figure}[h]
  \centering
  \begin{minipage}[b]{0.49\linewidth}
   \centering
   \includegraphics[width=\textwidth]{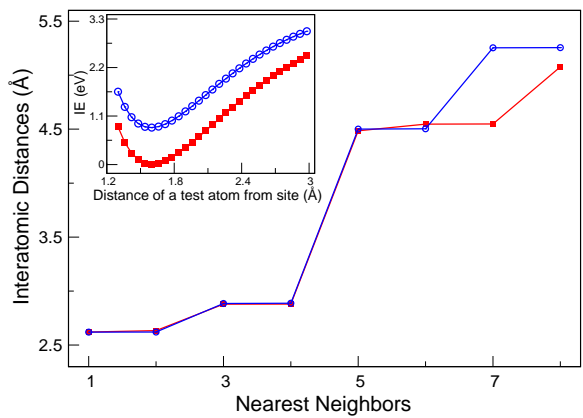}
  \end{minipage}
   \caption{The figure shows nearest neighbor distribution for two surface atoms of Al$_{9}$ cluster. The inset
shows distance dependent interaction energy for these two atoms. Two sites (or surface atoms) have 
identical interaction energy if and only if they have identical chemical environment defined in terms of 
all interatomic distances.}
  \label{fig5}
\end{figure}

It has been also observed that two sites result into identical interaction energy pattern, if and only if `all'
the interatomic distances are identical. For example, in the case of Al$_{9}$, for two atoms, their first 6 
nearest neighbors are at identical distances  and the last two distances differ as shown in Fig.\ \ref{fig5}.
However, it has resulted into two distinct interaction energy patterns for the respective atoms as shown 
in inset of Fig. \ref{fig5}.
This one to one correlation between the nearest neighbor distances and 
interaction energy pattern is observed in all the Al clusters that we have studied with size ranging
from 5 to 80. The same trend was observed when H atom was replaced with N atom, and N$_{2}$, O$_{2}$, and CO molecules.
Also, when the cluster is replaced with that of other elements like Na$_{10}$, and for bimetallic cluster like Al$_{6}$Ga$_{6}$, 
this correlation holds. 
This strong correlation between identical adsorption sites and identical interaction energy can be understood from
the perspective of Hohenberg-Kohn's first theorem. HK's first theorem describes the one to one correspondence between
external potential and charge density and hence energy (a functional of charge density). Identical sites are the ones
that have same relative distribution of atoms in the cluster i.e. identical nearest neighbor distribution.
Which in turn results into identical external potential when a test atom is placed at appropriate position (as described in computational details)
and hence, identical interaction energy.

Establishing a structure-property relation to understand reactivity problems better will greatly
reduce DFT based computational efforts. But, coming up with a set of descriptors that have characteristics like
transferability, universality, potential to capture accurate trends and yet be simple is still an
ongoing area of research. 
And so, it will be interesting to test if this one-to-one correlation between nearest neighbor distribution and interaction energy could be exploited  by employing 
data driven models at a minimal computational cost.
Since in this size regime, properties of clusters vary substantially by addition/removal of just an atom, it 
becomes important to study the interaction of clusters as a function of size.
The size and site specific interaction energy data generated through our extensive DFT computations was used to train the ML model. 
The data was fed to GBR to predict the interaction energies for all unique adsorption sites of the clusters between 
5 to 20 and few selected larger clusters (25, 36, 42, 55, 67, and 75). This is a balanced mixture of ordered and disordered clusters.
Taking a hint from the DFT investigations,
descriptors that captured this structure-property relation were designed. While modelling the interaction of clusters
with adsorbate the (dis)similarity between two adsorption sites had to be captured. And hence, nearest neighbor
distribution as seen by the adsorbate was the logical choice of descriptors. The chosen set of descriptors did
not represent any homometric pairs as the cluster geometries were fixed, while only the distance between adsorbate
and cluster varied. For homogeneous clusters, only distances were used as descriptors while 
nuclear charge was also included for bimetallic cluster.
\begingroup
\setlength{\tabcolsep}{4.5pt}
\setlength{\extrarowheight}{0.3cm}
\begin{table}[h]
\noindent\makebox[\textwidth][c]{
\begin{minipage}[c]{\textwidth}
\centering
\begin{tabular}{@{}c c c c c c c c c c c c c@{} }
\toprule
Cluster & \multicolumn{12}{c@{}}{AME as a function of (nearest neighbors as) descriptors}\\
size& nn2 & nn3 & nn4 & nn5 & nn6 & nn7 & nn8 & nn9 & nn10 & nn11 & nn12 & nn13\\
\midrule
5 & 0.085 & 0.084 & 0.050 & 0.050 &  &  &  &  &  &  &  &  \\

6 & 0.034 & 0.035 & 0.035 & 0.034 & 0.032 &  &  &  &  &  &  &  \\

7 & 0.179 & 0.180 & 0.178 & 0.059 & 0.051 & 0.051 &  &  &  &  &  &  \\

8 & 0.057 & 0.038 & 0.035 & 0.029 & 0.028 & 0.028 & 0.028 &  &  &  &  &  \\

9 & 0.226 & 0.228 & 0.219 & 0.133 & 0.127 & 0.122 & 0.055 & 0.054 &  &  &  &  \\

10 & 0.051 & 0.053 & 0.051 & 0.052 & 0.051 & 0.044 & 0.043 & 0.043 & 0.044 &  &  &  \\

11 & 0.281 & 0.268 & 0.194 & 0.169 & 0.114 & 0.096 & 0.096 & 0.079 & 0.060 & 0.060 &  &  \\

12 & 0.081 & 0.069 & 0.067 & 0.063 & 0.055 & 0.041 & 0.041 & 0.039 & 0.038 & 0.038 & 0.032 &  \\

13 & 0.131 & 0.132 & 0.130 & 0.082 & 0.083 & 0.086 & 0.026 & 0.026 & 0.026 & 0.025 & 0.024 & 0.021 \\
\bottomrule
\end{tabular}
\caption{Absolute Mean Errors (AME) as a function of descriptors (interatomic distances (nn)) are shown in the table for various clusters. As seen from the table, increasing representation of the system results into improved accuracy.
The overall error is 0.05 eV.}
\label{tab1}
\end{minipage}
}
\end{table}
\endgroup

A trend of reducing prediction errors with increasing system representation was seen for all the clusters that we 
had studied. For any surface site of an \textrm{n} atom cluster, there will be \textrm{n} distances 
as descriptors. These descriptors were distances from adsorbate to all the atoms in a cluster and hence invariant to rotation, translation, and permutation of the system. 
Since, the cluster geometries are always fixed for any given size, \textrm{n} distances are enough to 
represent the entire system and hence the external potential. The model was trained each time by gradually including more descriptors i.e. distances. 
In Tab.\ \ref{tab1} we list the variation in AME as a function of increasing number of 
descriptors for smaller cluster sizes. The variation in AME is correlated with interatomic distances from the H 
atom. We will discuss this further by closely analyzing the specific case of Al$_{13}$.
\begin{figure}[h]
  \centering
  \begin{minipage}[b]{0.49\linewidth}
   \centering 
   \includegraphics[width=\textwidth]{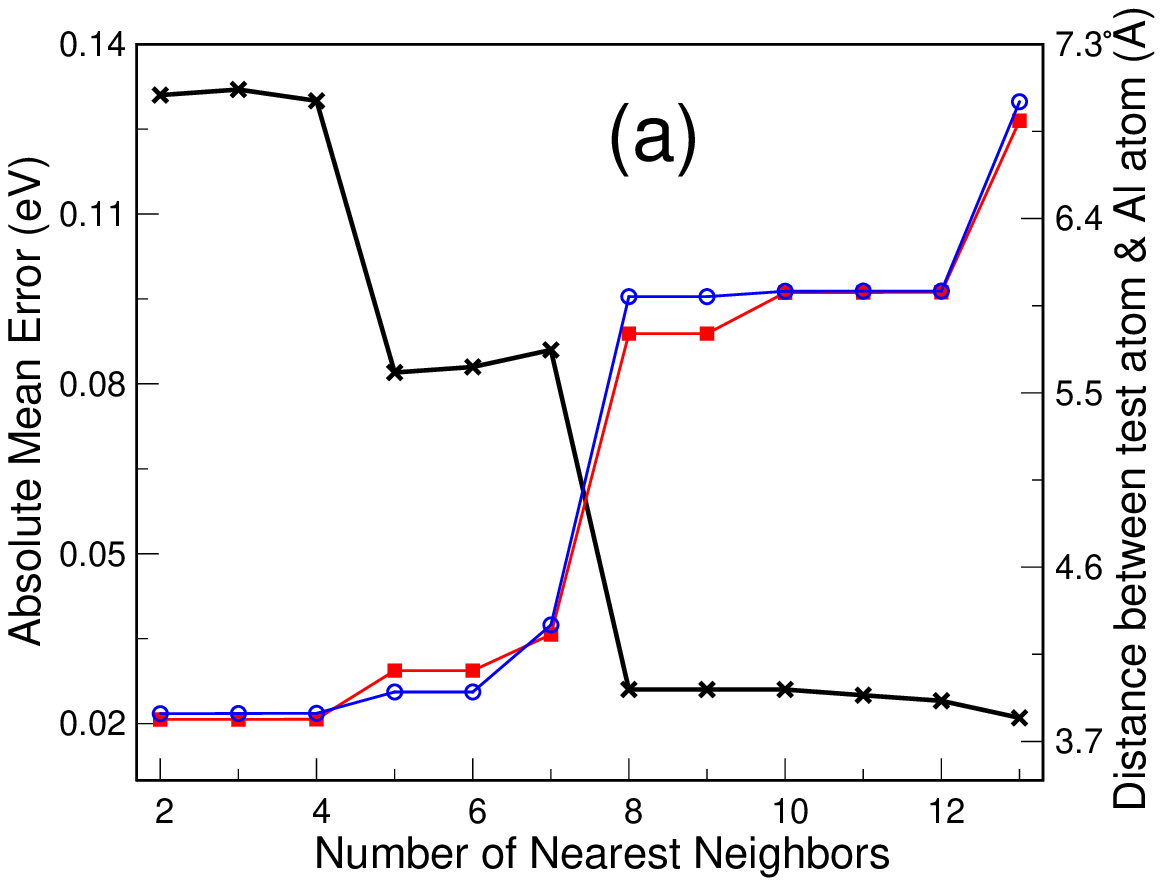}
  \end{minipage}
  \begin{minipage}[b]{0.49\linewidth}
   \centering 
   \includegraphics[width=\textwidth]{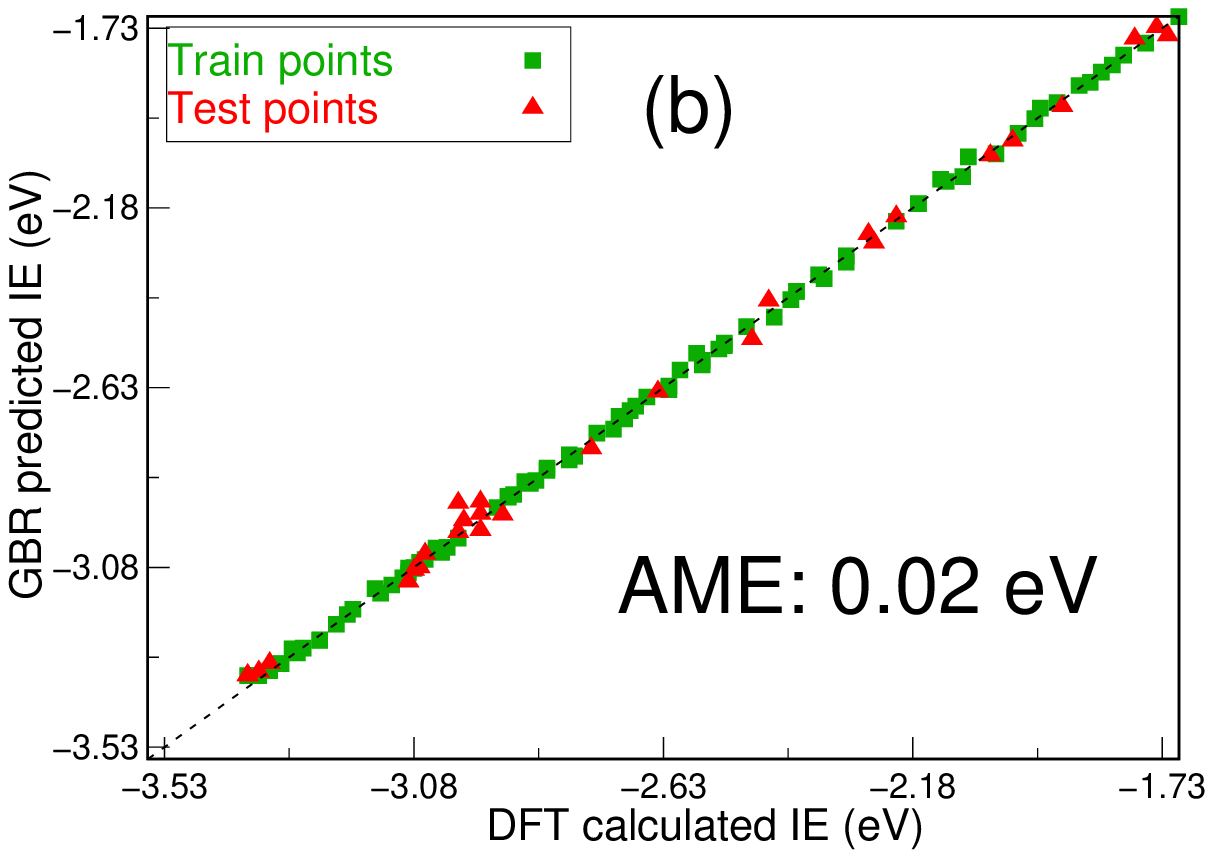}
  \end{minipage}
   \caption{(a) shows the decrease of errors (AME) with increasing number of descriptors for the Al$_{13}$ cluster plotted on y1 axis.
We have used the distance between H atom and surface atoms as descriptors which is plotted on y2 axis.
It is seen in figure that as the two groups in Al$_{13}$ are distinguished in NN distribution, the error reduces.
(b) shows absolute mean error in interaction energy prediction by ML model with reference to DFT.}
  \label{fig6}
\end{figure}
In Fig.\ \ref{fig6}-(a) we have plotted the AME as a function of number of descriptors (number of NN distances
from the surface site) used to fit the model for predicting interaction energies for Al$_{13}$. 
As noted earlier, Al$_{13}$ has only two types of atoms. The difference between these two types of atoms in their 
nearest neighbor distribution is picked up in the ML model. And hence we observed improvement in AME at distances 
where these two groups differ from each other i.e. AME reduced from 0.13 eV to 0.08 eV with descriptors up to nn4 
versus nn5. A similar jump (decrease) in AME was observed when nn8 was also included as shown in Tab.\ \ref{tab1}. 
nn8 is the point at which the two classes further separated. In Fig. \ \ref{fig6}-(b) ML predicted energies for 
Al$_{13}$ are plotted against DFT calculated energies. The AME in this specific case is 0.02 eV. 
\begin{figure}[h]
  \centering
  \begin{minipage}[b]{0.32\linewidth}
   \centering
   \includegraphics[width=\textwidth]{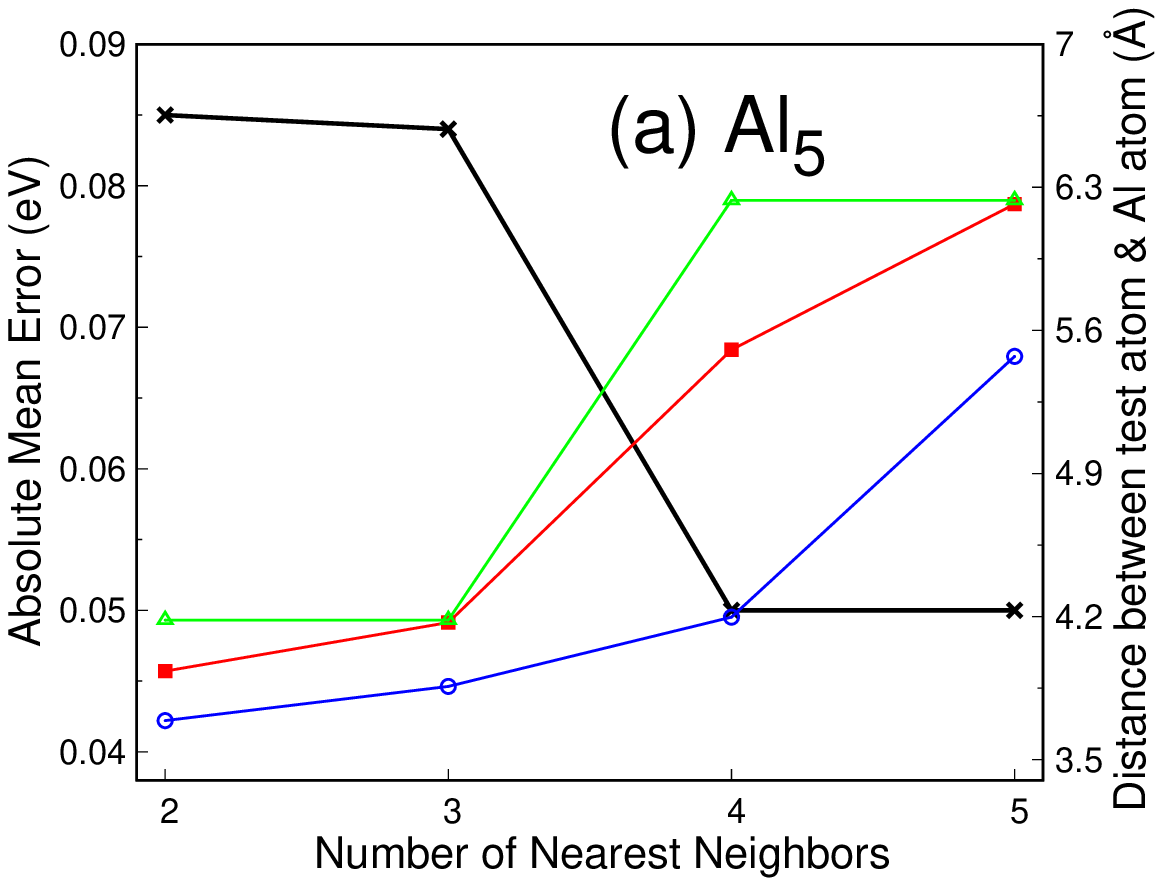}
  \end{minipage}
  \begin{minipage}[b]{0.32\linewidth}
   \centering
   \includegraphics[width=\textwidth]{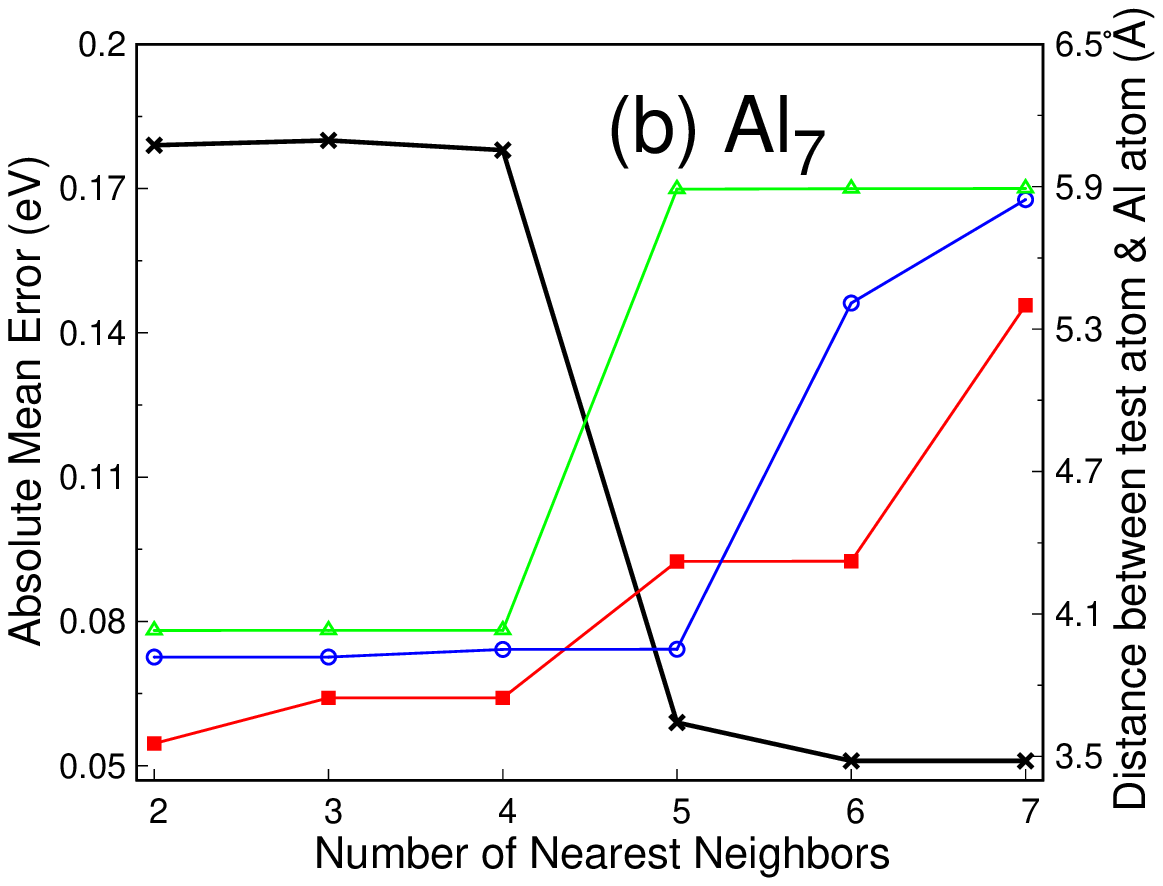}
  \end{minipage}
  \begin{minipage}[b]{0.32\linewidth}
   \centering
   \includegraphics[width=\textwidth]{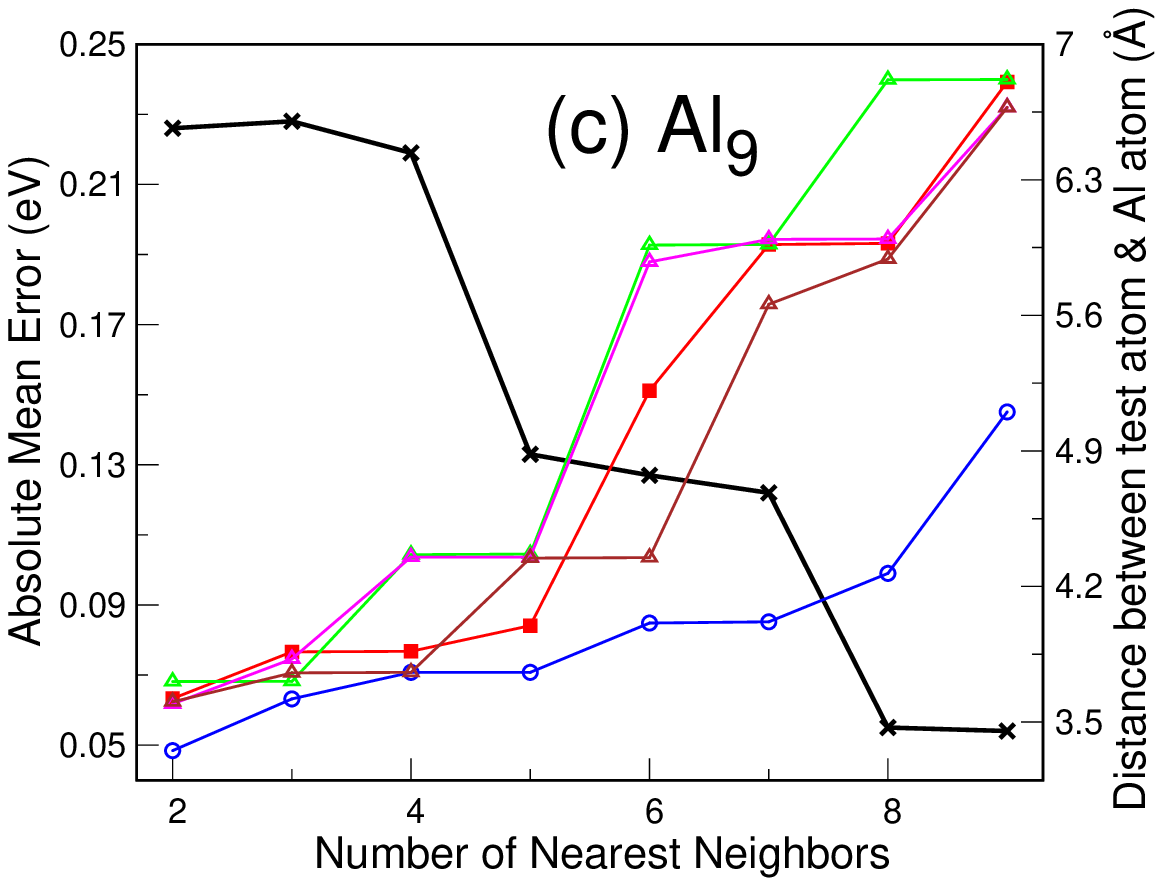}
  \end{minipage}
   \caption{Figure shows the decrease of errors (AME) with increasing number of descriptors for the clusters plotted on y1 axis.
We have used the distance between H atom and surface atoms as descriptors which is plotted on y2 axis.
}
  \label{fig7} 
\end{figure} 
This correlation between reduction in AME and variation in the nearest neighbor distribution was also 
observed for Al$_{5}$, Al$_{7}$, and Al$_{9}$ clusters (see Tab.\ \ref{tab1}). The nearest neighbor distribution 
and variation in AME plots for the above mentioned cases are shown in Fig.\ \ref{fig7}.
Overall, the machine learning model has picked up the underlying correlation between nearest neighbor 
distances and interaction energy. As has been discussed, by means of distances as descriptors, we are providing
information about the external potential and thus catching the essence of Hohenberg-Kohn's first theorem. 

\begingroup
\setlength{\extrarowheight}{0.3cm}
\setlength{\tabcolsep}{7pt}
\begin{table}[h]
\noindent\makebox[\textwidth][c]{
\begin{minipage}[c]{\textwidth}
\centering
\begin{tabular}{@{}c  c c c c c c c c c  @{} }
\toprule
Cluster & \multicolumn{8}{c@{}}{AME as a function of nearest neighbors as descriptors}\\
size & nn2 & nn3 & nn4 & nn5 & nn10 & nn 25\% & nn 50\%  & nn 100\% \\
\midrule
25 & 0.137 & 0.111 & 0.113 & 0.107 & 0.071 & 0.095(6) & 0.067(13) & 0.068(24) \\

36 & 0.063 & 0.050 & 0.054 & 0.056 & 0.044 & 0.045(9) & 0.044(18) & 0.030 (35) \\

42 & 0.083 & 0.084 & 0.077 & 0.074 & 0.076 & 0.076(11) & 0.069(21) & 0.059(41) \\

55 & 0.125 & 0.110 & 0.104 & 0.100 & 0.090 & 0.090(14) & 0.088(28) & 0.086(54) \\

67 & 0.083 & 0.077 & 0.072 & 0.066 & 0.048 & 0.048(17) & 0.045(35) & 0.049(66) \\

75 & 0.119 & 0.125 & 0.100 & 0.081 & 0.081 & 0.080 (18) & 0.071(37) & 0.060(74) \\

\bottomrule
\end{tabular}
\caption {Absolute mean errors (AME) as function of descriptors (interatomic distances (nn))
are shown in the table for larger clusters.  The last three columns present 25\%, 50\%, and 100\%
of the system representation respectively.  The numbers in the brackets indicate the number of 
interatomic distances used to predict the interaction energy.}
\label{tab2}
\end{minipage}
}
\end{table}
\endgroup
This correlation between nearest neighbor and variation in AME is strikingly evident and easy to capture in smaller clusters. While it is
not so clear in the case of larger cluster due to increased complexity of the systems.
And this is what reflects into the AME as a function of descriptors as shown in Tab.\ \ref{tab2}.
It was observed that the variation in AME was inconsistent when descriptors up to nn10 versus all distances 
(nn100\%) were used. For example, in the case of Al$_{36}$ and Al$_{75}$, reduction in AME was more than
25\% for each of them as shown in Tab. \ \ref{tab2}. It must be noted that Al$_{36}$ (see Fig.\ \ref{fig3}-(a)) and 
Al$_{75}$ (see Fig.\ \ref{fig4}-(c)) are highly symmetric clusters. Whereas for asymmetric clusters like Al$_{25}$,  
Al$_{55}$, and Al$_{67}$ the reduction in errors were less than 5\% as evident from Tab.\ \ref{tab2}. 
But for another asymmetric cluster, Al$_{42}$, the reduction is much larger i.e. about 20\% which  
is similar to that of symmetric clusters. Thus generalization of results becomes difficult for larger 
clusters. Nonetheless, even for larger clusters the one to one correlation between reduction in AME and 
increasing system representation still holds. The overall AME reported in our work is $\approx$ 0.05 eV.

\begin{figure}
  \centering
  \begin{minipage}[b]{0.35\linewidth}
   \centering
   \textbf{(a) DFT calculated PES}\par\medskip
   \includegraphics[width=\textwidth]{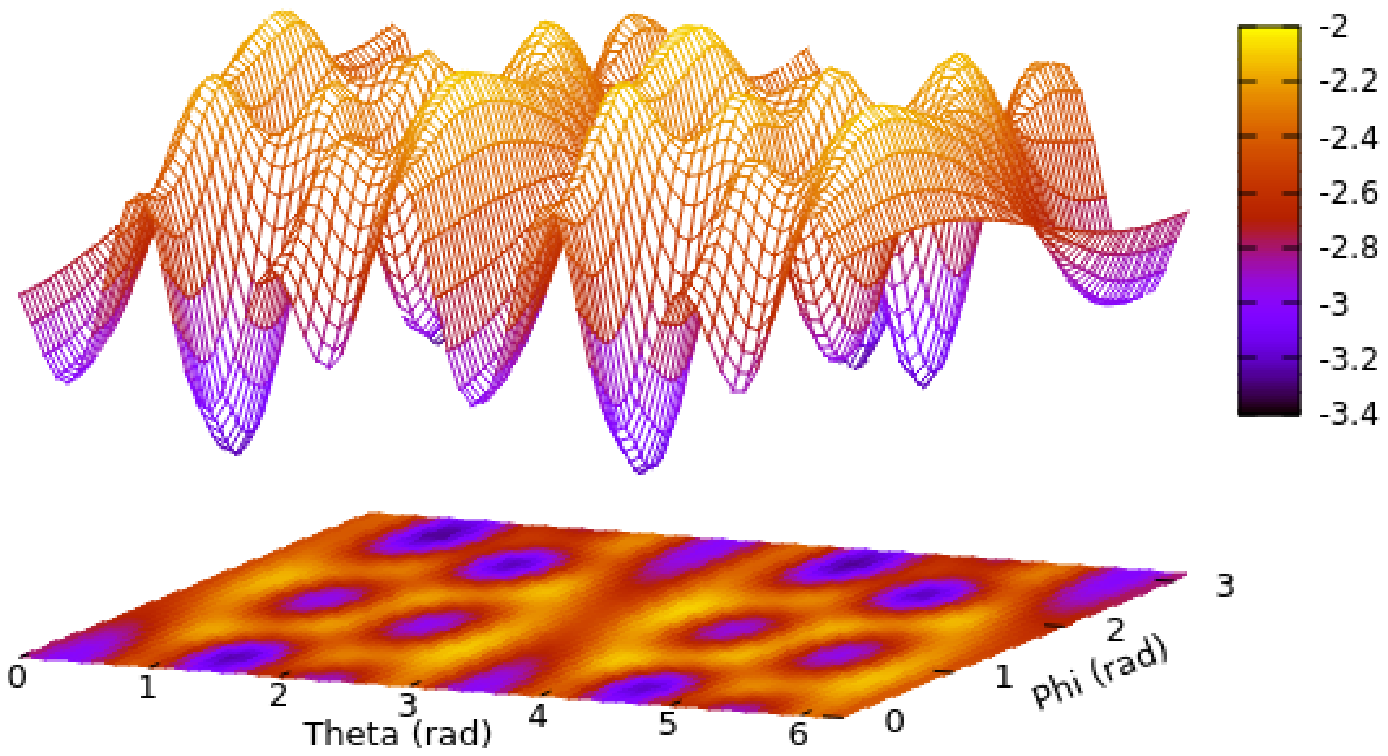}
  \end{minipage}
  \begin{minipage}[b]{0.26\linewidth}
   \centering
   \textbf{(b) Schematic of Al$_{13}$ with sample points}\par\medskip
   \includegraphics[width=\textwidth]{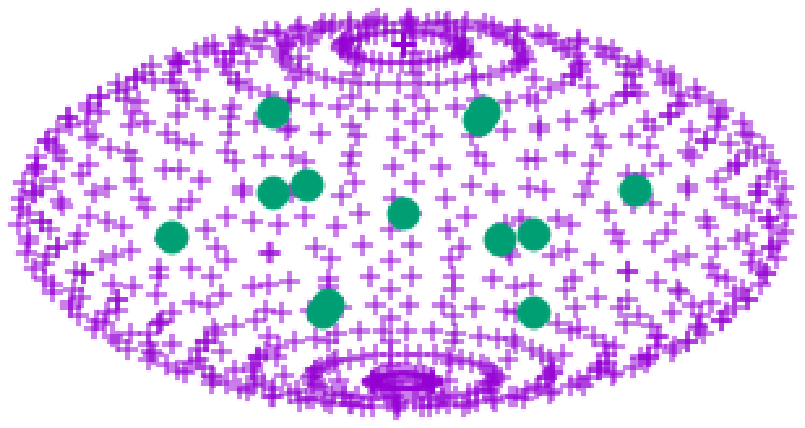}
  \end{minipage}
  \begin{minipage}[b]{0.35\linewidth}
   \centering
   \textbf{(c) ML predicted PES}\par\medskip
   \includegraphics[width=\textwidth]{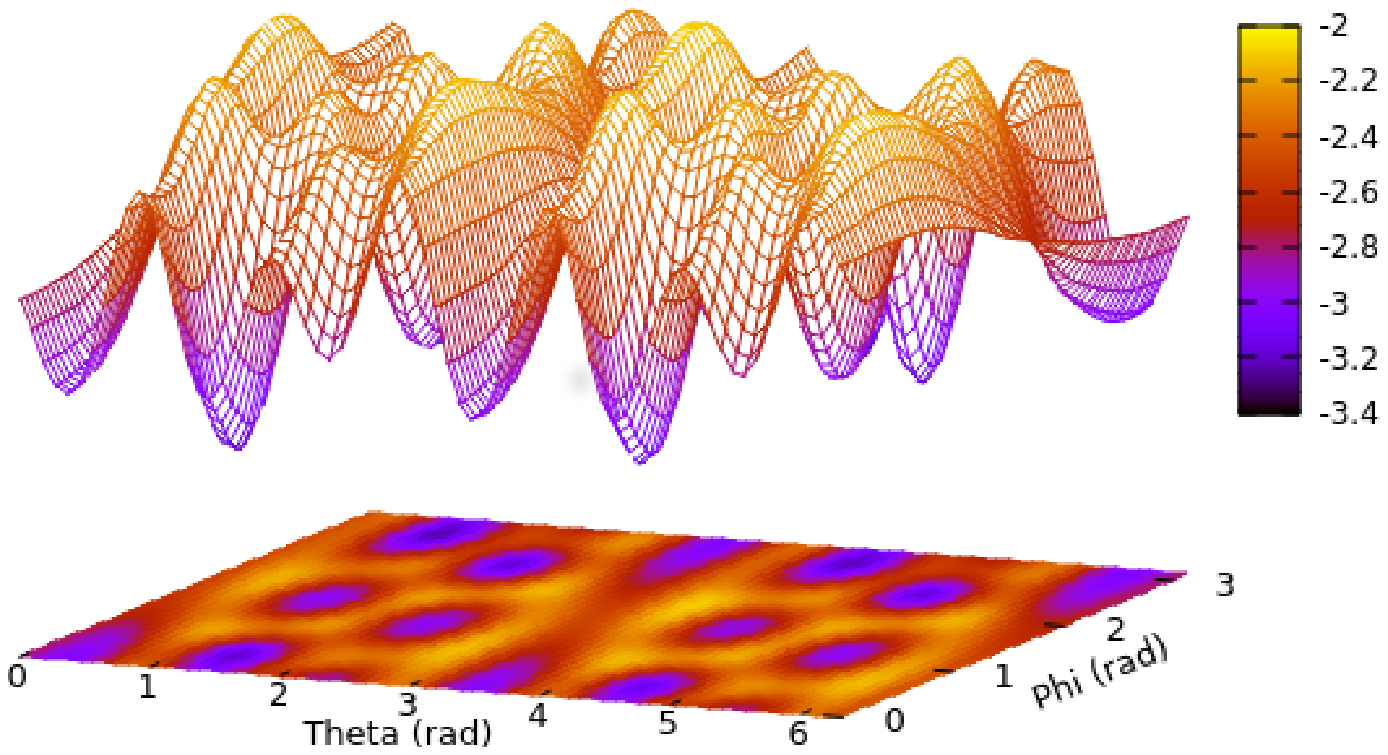}
  \end{minipage}
   \caption{(a) PES computed through DFT calculations. (b) cartoon of Al$_{13}$ cluster and points selected on a sphere 
to compute interaction of adsorbate with cluster. (c) ML generated PES. It is evident that our model
has picked up the variation in PES quite faithfully.
}
  \label{fig8} 
\end{figure} 

Since the line of search for all the results discussed above was restricted along the radial vector, to model
a real situation wherein an adsorbate can approach the cluster from any direction, all possible directions 
had to be scanned. 
The one to one correlation between identical sites and identical interaction would be difficult to quantify for this situation,
as now the adsorbate was not placed only at on-top sites. Nonetheless, the same recipe of descriptors
was still legitimate as the distances taken were from the adsorbate to the atoms representing the external potential. 
And so, the same set of descriptors would capture the change in chemical environment as seen by an incoming
adsorbate. 
Thus, a logical proposition is, the potential energy surface (PES) of an adsorbate in the
vicinity of a cluster could be explored with our model. To validate this, we scanned PES at 800 different points
(in all possible directions) around a cluster and is shown in Fig\ \ref{fig8}.
We computed the interaction energy of H atom for these randomly selected 800 points
on a sphere that enclosed the Al$_{13}$ cluster at its center (see Fig.\ \ref{fig8}-(b)). The distance of H
atom from the closest surface site of cluster varies between 1.60 \AA~ to 2.69 \AA. This result is particularly important because
through this we could predict interaction energy at any point on the PES of the cluster-adsorbate system
with AME as low as 0.04 eV. 
\begin{figure}[h]
  \centering
  \begin{minipage}[b]{0.49\linewidth}
   \centering
   \includegraphics[width=\textwidth]{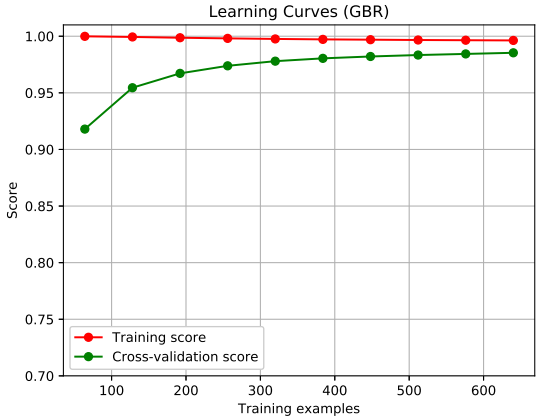}
  \end{minipage}
   \caption{It is evident from the learning curve that the training score is maximum and cross validation score increases 
gradually with increasing data points.}
  \label{fig9}
\end{figure}
The success of ML model, in this case, is a proof of concept that nearest neighbor distances
are the correct choice of descriptors. 
It was also found that with only 400 points on the sphere we could achieve same level of accuracy as with
800 points as shown in learning curve Fig.\ \ref{fig9}. Further as seen from the figure, our model has picked up variation in PES quite faithfully. The minima
on the PES represents on-top position which is the most favorable position for the H atom on this surface. 
Whereas the maxima is the least favorable position and turns out to be a bridge position for the H atom. 

To further validate our model we tested it on other clusters like Na$_{10}$ and Al$_{6}$Ga$_{6}$. It was observed that 
for H atom adsorbed on a highly asymmetric Na$_{10}$ cluster, our ML model with the same recipe of descriptors predicted 
the IE with AME $\approx$ 0.038 eV. 
To demonstrate the universality of our work, calculations performed with different adsorbing species on 
Al clusters are noted below.
When a single N atom was adsorbed at on-top positions on Al$_{13}$ cluster, not only the one to one correlation was observed again
but also the AME from ML model was 0.06 eV, i.e. in same range as our previous results.
Further the errors for prediction of IE using the same ML model when molecules like N$_{2}$, O$_{2}$, and CO were adsorbed on Al$_{12}$ turned out to be,
0.045 eV, 0.049 eV, and 0.042 eV respectively.
Finally, we also tested the validity of our ML model on bimetallic cluster Al$_{6}$Ga$_{6}$. AME for an H atom 
adsorbed on top of this cluster turned out to be 0.09 eV. This error was obtained based on only structural representation 
of the cluster. With the inclusion of nuclear charge/ionic radii of both the elements of the cluster, viz. Al and Ga, 
in the descriptor set, the AME got down to 0.058 eV. 

In a nutshell, the descriptors used to train the ML model are as simple as distances between adsorbate and the surface 
atoms correlating the structure and activity between the two. Successful prediction of interaction energy by
means of descriptors that systematically represent the external potential catches the essence of Hohenberg-Kohn's
first theorem. Our approach differs from the previous work in a few key ways: 1. Descriptors chosen were such so as to model the 
interaction of a cluster and an incoming adsorbate. And hence, the chemical environment that an adsorbate experienced was explored,
2. The descriptors used did not represent any homometric pairs as only unique adsorption sites for fixed cluster geometries
were used and finally 3. The ML model was trained on purely the structural representation of the cluster.
\section{Conclusion}
To summarize we have combined DFT with ML to understand the interaction between an adsorbate and clusters.
The key results of the present work are as follows:
1. Our extensive DFT calculations establish a one-to-one correlation between the nearest neighbor distances 
and interaction energy for small Al clusters. The results, as demonstrated, are generic and applicable to all the clusters and different adsorbates.
2. We employ the GBR model to predict the site specific interaction energies by using `only' interatomic 
distances as descriptors. The absolute mean errors are about 0.05 eV. With this, we also demonstrate that our ML
algorithm picks up the one-to-one correlation between the nearest neighbor distribution and the site specific 
interaction energies and hence essence of Hohenberg-Kohn's first theorem.
3. We reproduce the PES for a test atom in the vicinity of the cluster by employing our ML model.
To get AME about as low as 0.04 eV we require only 400 single point calculations, which demonstrates that 
we could circumvent the compute intensive DFT by employing this model. 4. Our descriptors are 
the interatomic distances, and hence the computational cost is negligible. 
In conclusion, we designed a set of descriptors that were as simple as nearest neighbor distances and yet the ones
that could accurately capture the structure-activity relation between the cluster and adsorbate. 

\section{Acknowledgements}
The authors thank Dr. Leelavati Narlikar for fruitful discussions.
CSIR-4PI is gratefully acknowledged for the computational facility. KJ acknowledges DST (EMR/2016/000591) 
for partial financial support. SM acknowledges UGC for research fellowship. SA acknowledges DST-INSPIRE 
for research fellowship.
\bibliography{Reference}

\providecommand{\latin}[1]{#1}
\providecommand*\mcitethebibliography{\thebibliography}
\csname @ifundefined\endcsname{endmcitethebibliography}
  {\let\endmcitethebibliography\endthebibliography}{}
\begin{mcitethebibliography}{83}
\providecommand*\natexlab[1]{#1}
\providecommand*\mciteSetBstSublistMode[1]{}
\providecommand*\mciteSetBstMaxWidthForm[2]{}
\providecommand*\mciteBstWouldAddEndPuncttrue
  {\def\EndOfBibitem{\unskip.}}
\providecommand*\mciteBstWouldAddEndPunctfalse
  {\let\EndOfBibitem\relax}
\providecommand*\mciteSetBstMidEndSepPunct[3]{}
\providecommand*\mciteSetBstSublistLabelBeginEnd[3]{}
\providecommand*\EndOfBibitem{}
\mciteSetBstSublistMode{f}
\mciteSetBstMaxWidthForm{subitem}{(\alph{mcitesubitemcount})}
\mciteSetBstSublistLabelBeginEnd
  {\mcitemaxwidthsubitemform\space}
  {\relax}
  {\relax}

\bibitem[Jortner(1992)]{jortner1992}
Jortner,~J. Cluster size effects. \emph{Zeitschrift f{\"u}r Physik D Atoms,
  Molecules and Clusters} \textbf{1992}, \emph{24}, 247--275\relax
\mciteBstWouldAddEndPuncttrue
\mciteSetBstMidEndSepPunct{\mcitedefaultmidpunct}
{\mcitedefaultendpunct}{\mcitedefaultseppunct}\relax
\EndOfBibitem
\bibitem[Sanchez \latin{et~al.}(1999)Sanchez, Abbet, Heiz, Schneider,
  Häkkinen, Barnett, and Landman]{AuLandman}
Sanchez,~A.; Abbet,~S.; Heiz,~U.; Schneider,~W.-D.; Häkkinen,~H.;
  Barnett,~R.~N.; Landman,~U. When Gold Is Not Noble: Nanoscale Gold Catalysts.
  \emph{J. Phys. Chem. A} \textbf{1999}, \emph{103}, 9573--9578\relax
\mciteBstWouldAddEndPuncttrue
\mciteSetBstMidEndSepPunct{\mcitedefaultmidpunct}
{\mcitedefaultendpunct}{\mcitedefaultseppunct}\relax
\EndOfBibitem
\bibitem[Jena and Castleman(2006)Jena, and Castleman]{PNAS2006}
Jena,~P.; Castleman,~A.~W. Clusters: A bridge across the disciplines of physics
  and chemistry. \emph{Proc. Natl. Acad. Sci.} \textbf{2006}, \emph{103},
  10560--10569\relax
\mciteBstWouldAddEndPuncttrue
\mciteSetBstMidEndSepPunct{\mcitedefaultmidpunct}
{\mcitedefaultendpunct}{\mcitedefaultseppunct}\relax
\EndOfBibitem
\bibitem[Castleman~Jr and Khanna(2009)Castleman~Jr, and Khanna]{castleman2009}
Castleman~Jr,~A.; Khanna,~S. Clusters, superatoms, and building blocks of new
  materials. \emph{J. Phys. Chem. C} \textbf{2009}, \emph{113},
  2664--2675\relax
\mciteBstWouldAddEndPuncttrue
\mciteSetBstMidEndSepPunct{\mcitedefaultmidpunct}
{\mcitedefaultendpunct}{\mcitedefaultseppunct}\relax
\EndOfBibitem
\bibitem[Alonso(2012)]{alonso2012structure}
Alonso,~J.~A. \emph{Structure and properties of atomic nanoclusters}; World
  Scientific, 2012\relax
\mciteBstWouldAddEndPuncttrue
\mciteSetBstMidEndSepPunct{\mcitedefaultmidpunct}
{\mcitedefaultendpunct}{\mcitedefaultseppunct}\relax
\EndOfBibitem
\bibitem[Berry and Smirnov(2013)Berry, and Smirnov]{berry2013bridging}
Berry,~R.~S.; Smirnov,~B.~M. Bridging the macro and micro. \emph{Chem. Phys.
  Lett} \textbf{2013}, \emph{573}, 1--4\relax
\mciteBstWouldAddEndPuncttrue
\mciteSetBstMidEndSepPunct{\mcitedefaultmidpunct}
{\mcitedefaultendpunct}{\mcitedefaultseppunct}\relax
\EndOfBibitem
\bibitem[Schmidt \latin{et~al.}(1998)Schmidt, Kusche, von Issendorff, and
  Haberland]{schmidt1998}
Schmidt,~M.; Kusche,~R.; von Issendorff,~B.; Haberland,~H. Irregular variations
  in the melting point of size-selected atomic clusters. \emph{Nature}
  \textbf{1998}, \emph{393}, 238\relax
\mciteBstWouldAddEndPuncttrue
\mciteSetBstMidEndSepPunct{\mcitedefaultmidpunct}
{\mcitedefaultendpunct}{\mcitedefaultseppunct}\relax
\EndOfBibitem
\bibitem[Breaux \latin{et~al.}(2004)Breaux, Hillman, Neal, Benirschke, and
  Jarrold]{jarrold2004}
Breaux,~G.~A.; Hillman,~D.~A.; Neal,~C.~M.; Benirschke,~R.~C.; Jarrold,~M.~F.
  Gallium Cluster “Magic Melters”. \emph{J. Am. Chem. Soc.} \textbf{2004},
  \emph{126}, 8628\relax
\mciteBstWouldAddEndPuncttrue
\mciteSetBstMidEndSepPunct{\mcitedefaultmidpunct}
{\mcitedefaultendpunct}{\mcitedefaultseppunct}\relax
\EndOfBibitem
\bibitem[Joshi \latin{et~al.}(2006)Joshi, Krishnamurty, and
  Kanhere]{kavita2006}
Joshi,~K.; Krishnamurty,~S.; Kanhere,~D. “Magic melters” have geometrical
  origin. \emph{Phys. Rev. Lett.} \textbf{2006}, \emph{96}, 135703\relax
\mciteBstWouldAddEndPuncttrue
\mciteSetBstMidEndSepPunct{\mcitedefaultmidpunct}
{\mcitedefaultendpunct}{\mcitedefaultseppunct}\relax
\EndOfBibitem
\bibitem[Berry and Smirnov(2013)Berry, and Smirnov]{berry2013}
Berry,~R.~S.; Smirnov,~B.~M. Configurational transitions in processes involving
  metal clusters. \emph{Phys. Rep} \textbf{2013}, \emph{527}, 205--250\relax
\mciteBstWouldAddEndPuncttrue
\mciteSetBstMidEndSepPunct{\mcitedefaultmidpunct}
{\mcitedefaultendpunct}{\mcitedefaultseppunct}\relax
\EndOfBibitem
\bibitem[Shvartsburg and Jarrold(1999)Shvartsburg, and Jarrold]{jarrold1999}
Shvartsburg,~A.~A.; Jarrold,~M.~F. Tin clusters adopt prolate geometries.
  \emph{Phys. Rev. A} \textbf{1999}, \emph{60}, 1235\relax
\mciteBstWouldAddEndPuncttrue
\mciteSetBstMidEndSepPunct{\mcitedefaultmidpunct}
{\mcitedefaultendpunct}{\mcitedefaultseppunct}\relax
\EndOfBibitem
\bibitem[Cox \latin{et~al.}(1988)Cox, Trevor, Whetten, and Kaldor]{cox1988}
Cox,~D.; Trevor,~D.; Whetten,~R.; Kaldor,~A. Aluminum clusters: ionization
  thresholds and reactivity toward deuterium, water, oxygen, methanol, methane,
  and carbon monoxide. \emph{J. Phys. Chem.} \textbf{1988}, \emph{92},
  421--429\relax
\mciteBstWouldAddEndPuncttrue
\mciteSetBstMidEndSepPunct{\mcitedefaultmidpunct}
{\mcitedefaultendpunct}{\mcitedefaultseppunct}\relax
\EndOfBibitem
\bibitem[Argo \latin{et~al.}(2002)Argo, Odzak, Lai, and Gates]{argo2002}
Argo,~A.; Odzak,~J.; Lai,~F.; Gates,~B. Observation of ligand effects during
  alkene hydrogenation catalysed by supported metal clusters. \emph{Nature}
  \textbf{2002}, \emph{415}, 623--626\relax
\mciteBstWouldAddEndPuncttrue
\mciteSetBstMidEndSepPunct{\mcitedefaultmidpunct}
{\mcitedefaultendpunct}{\mcitedefaultseppunct}\relax
\EndOfBibitem
\bibitem[Fu \latin{et~al.}(2003)Fu, Saltsburg, and
  Flytzani-Stephanopoulos]{fu2003}
Fu,~Q.; Saltsburg,~H.; Flytzani-Stephanopoulos,~M. Active nonmetallic Au and Pt
  species on ceria-based water-gas shift catalysts. \emph{Science}
  \textbf{2003}, \emph{301}, 935--938\relax
\mciteBstWouldAddEndPuncttrue
\mciteSetBstMidEndSepPunct{\mcitedefaultmidpunct}
{\mcitedefaultendpunct}{\mcitedefaultseppunct}\relax
\EndOfBibitem
\bibitem[Campbell(2004)]{campbell2004}
Campbell,~C.~T. The active site in nanoparticle gold catalysis. \emph{Science}
  \textbf{2004}, \emph{306}, 234--235\relax
\mciteBstWouldAddEndPuncttrue
\mciteSetBstMidEndSepPunct{\mcitedefaultmidpunct}
{\mcitedefaultendpunct}{\mcitedefaultseppunct}\relax
\EndOfBibitem
\bibitem[Chen and Goodman(2004)Chen, and Goodman]{chen2004}
Chen,~M.; Goodman,~D. The structure of catalytically active gold on titania.
  \emph{Science} \textbf{2004}, \emph{306}, 252--255\relax
\mciteBstWouldAddEndPuncttrue
\mciteSetBstMidEndSepPunct{\mcitedefaultmidpunct}
{\mcitedefaultendpunct}{\mcitedefaultseppunct}\relax
\EndOfBibitem
\bibitem[Lemire \latin{et~al.}(2004)Lemire, Meyer, Shaikhutdinov, and
  Freund]{lemire2004}
Lemire,~C.; Meyer,~R.; Shaikhutdinov,~S.; Freund,~H.-J. Do quantum size effects
  control CO adsorption on gold nanoparticles? \emph{Angew. Chem.}
  \textbf{2004}, \emph{43}, 118--121\relax
\mciteBstWouldAddEndPuncttrue
\mciteSetBstMidEndSepPunct{\mcitedefaultmidpunct}
{\mcitedefaultendpunct}{\mcitedefaultseppunct}\relax
\EndOfBibitem
\bibitem[Wei and Iglesia(2004)Wei, and Iglesia]{wei2004}
Wei,~J.; Iglesia,~E. Mechanism and site requirements for activation and
  chemical conversion of methane on supported Pt clusters and turnover rate
  comparisons among noble metals. \emph{J. Phys. Chem. B} \textbf{2004},
  \emph{108}, 4094--4103\relax
\mciteBstWouldAddEndPuncttrue
\mciteSetBstMidEndSepPunct{\mcitedefaultmidpunct}
{\mcitedefaultendpunct}{\mcitedefaultseppunct}\relax
\EndOfBibitem
\bibitem[Vajda \latin{et~al.}(2009)Vajda, Pellin, Greeley, Marshall, Curtiss,
  Ballentine, Elam, Catillon-Mucherie, Redfern, Mehmood, and Zapol]{vajdaPt}
Vajda,~S.; Pellin,~M.~J.; Greeley,~J.~P.; Marshall,~C.~L.; Curtiss,~L.~A.;
  Ballentine,~G.~A.; Elam,~J.~W.; Catillon-Mucherie,~S.; Redfern,~P.~C.;
  Mehmood,~F. \latin{et~al.}  Subnanometre platinum clusters as highly active
  and selective catalysts for the oxidative dehydrogenation of propane.
  \emph{Nat. Mater.} \textbf{2009}, \emph{8}, 213--216\relax
\mciteBstWouldAddEndPuncttrue
\mciteSetBstMidEndSepPunct{\mcitedefaultmidpunct}
{\mcitedefaultendpunct}{\mcitedefaultseppunct}\relax
\EndOfBibitem
\bibitem[Heiz \latin{et~al.}(1998)Heiz, Vanolli, Sanchez, and
  Schneider]{heiz1998}
Heiz,~U.; Vanolli,~F.; Sanchez,~A.; Schneider,~W.-D. Size-dependent molecular
  dissociation on mass-selected, supported metal clusters. \emph{J. Am. Chem.
  Soc.} \textbf{1998}, \emph{120}, 9668--9671\relax
\mciteBstWouldAddEndPuncttrue
\mciteSetBstMidEndSepPunct{\mcitedefaultmidpunct}
{\mcitedefaultendpunct}{\mcitedefaultseppunct}\relax
\EndOfBibitem
\bibitem[Wallace and Whetten(2000)Wallace, and Whetten]{wallace2000}
Wallace,~W.~T.; Whetten,~R.~L. Carbon monoxide adsorption on selected gold
  clusters: highly size-dependent activity and saturation compositions.
  \emph{J. Phys. Chem. B} \textbf{2000}, \emph{104}, 10964--10968\relax
\mciteBstWouldAddEndPuncttrue
\mciteSetBstMidEndSepPunct{\mcitedefaultmidpunct}
{\mcitedefaultendpunct}{\mcitedefaultseppunct}\relax
\EndOfBibitem
\bibitem[Cao \latin{et~al.}(2009)Cao, Starace, Judd, and Jarrold]{jarroldjacs}
Cao,~B.; Starace,~A.~K.; Judd,~O.~H.; Jarrold,~M.~F. Melting Dramatically
  Enhances the Reactivity of Aluminum Nanoclusters. \emph{J. Am. Chem. Soc.}
  \textbf{2009}, \emph{131}, 2446--2447\relax
\mciteBstWouldAddEndPuncttrue
\mciteSetBstMidEndSepPunct{\mcitedefaultmidpunct}
{\mcitedefaultendpunct}{\mcitedefaultseppunct}\relax
\EndOfBibitem
\bibitem[Roach \latin{et~al.}(2009)Roach, Woodward, Castleman, Reber, and
  Khanna]{roach2009}
Roach,~P.~J.; Woodward,~W.~H.; Castleman,~A.; Reber,~A.~C.; Khanna,~S.~N.
  Complementary active sites cause size-selective reactivity of aluminum
  cluster anions with water. \emph{Science} \textbf{2009}, \emph{323},
  492--495\relax
\mciteBstWouldAddEndPuncttrue
\mciteSetBstMidEndSepPunct{\mcitedefaultmidpunct}
{\mcitedefaultendpunct}{\mcitedefaultseppunct}\relax
\EndOfBibitem
\bibitem[Reber \latin{et~al.}(2010)Reber, Khanna, Roach, Woodward, and
  Castleman~Jr]{reber2010}
Reber,~A.~C.; Khanna,~S.~N.; Roach,~P.~J.; Woodward,~W.~H.; Castleman~Jr,~A.
  Reactivity of aluminum cluster anions with water: origins of reactivity and
  mechanisms for H2 release. \emph{J. Phys. Chem. A} \textbf{2010}, \emph{114},
  6071--6081\relax
\mciteBstWouldAddEndPuncttrue
\mciteSetBstMidEndSepPunct{\mcitedefaultmidpunct}
{\mcitedefaultendpunct}{\mcitedefaultseppunct}\relax
\EndOfBibitem
\bibitem[Kulkarni \latin{et~al.}(2011)Kulkarni, Krishnamurty, and
  Pal]{kulkarni2011}
Kulkarni,~B.~S.; Krishnamurty,~S.; Pal,~S. Size- and Shape-Sensitive Reactivity
  Behavior of Aln (n = 2–5, 13, 30, and 100) Clusters Toward the N2 Molecule:
  A First-Principles Investigation. \emph{J. Phys. Chem. C} \textbf{2011},
  \emph{115}, 14615--14623\relax
\mciteBstWouldAddEndPuncttrue
\mciteSetBstMidEndSepPunct{\mcitedefaultmidpunct}
{\mcitedefaultendpunct}{\mcitedefaultseppunct}\relax
\EndOfBibitem
\bibitem[Yin and Bernstein(2012)Yin, and Bernstein]{yin2012gas}
Yin,~S.; Bernstein,~E.~R. Gas phase chemistry of neutral metal clusters:
  distribution, reactivity and catalysis. \emph{Int. J. Mass Spectrom.}
  \textbf{2012}, \emph{321}, 49--65\relax
\mciteBstWouldAddEndPuncttrue
\mciteSetBstMidEndSepPunct{\mcitedefaultmidpunct}
{\mcitedefaultendpunct}{\mcitedefaultseppunct}\relax
\EndOfBibitem
\bibitem[Luo \latin{et~al.}(2016)Luo, Castleman~Jr, and Khanna]{review2016}
Luo,~Z.; Castleman~Jr,~A.; Khanna,~S.~N. Reactivity of metal clusters.
  \emph{Chem. Rev.} \textbf{2016}, \emph{116}, 14456--14492\relax
\mciteBstWouldAddEndPuncttrue
\mciteSetBstMidEndSepPunct{\mcitedefaultmidpunct}
{\mcitedefaultendpunct}{\mcitedefaultseppunct}\relax
\EndOfBibitem
\bibitem[Das \latin{et~al.}(2013)Das, Pal, and Krishnamurty]{pal2013}
Das,~S.; Pal,~S.; Krishnamurty,~S. Understanding the site selectivity in
  small-sized neutral and charged Al$_{n}$ (4 $\leq$ n $\leq$ 7) clusters using
  density functional theory based reactivity descriptors: a validation study on
  water molecule adsorption. \emph{J. Phys. Chem. A} \textbf{2013}, \emph{117},
  8691--8702\relax
\mciteBstWouldAddEndPuncttrue
\mciteSetBstMidEndSepPunct{\mcitedefaultmidpunct}
{\mcitedefaultendpunct}{\mcitedefaultseppunct}\relax
\EndOfBibitem
\bibitem[Luo \latin{et~al.}(2013)Luo, Smith, Berkdemir, and
  Castleman]{castleman2013}
Luo,~Z.; Smith,~J.~C.; Berkdemir,~C.; Castleman,~A. Gas-phase reactivity of
  aluminum cluster anions with ethanethiol: Carbon--sulfur bond activation.
  \emph{Chem. Phys. Lett.} \textbf{2013}, \emph{590}, 63--68\relax
\mciteBstWouldAddEndPuncttrue
\mciteSetBstMidEndSepPunct{\mcitedefaultmidpunct}
{\mcitedefaultendpunct}{\mcitedefaultseppunct}\relax
\EndOfBibitem
\bibitem[Mueller \latin{et~al.}(2016)Mueller, Kusne, and Ramprasad]{MLinmatsci}
Mueller,~T.; Kusne,~A.~G.; Ramprasad,~R. Machine learning in materials science:
  Recent progress and emerging applications. \emph{Reviews in Computational
  Chemistry} \textbf{2016}, \emph{29}, 186--273\relax
\mciteBstWouldAddEndPuncttrue
\mciteSetBstMidEndSepPunct{\mcitedefaultmidpunct}
{\mcitedefaultendpunct}{\mcitedefaultseppunct}\relax
\EndOfBibitem
\bibitem[Butler \latin{et~al.}(2018)Butler, Davies, Cartwright, Isayev, and
  Walsh]{butler2018machine}
Butler,~K.~T.; Davies,~D.~W.; Cartwright,~H.; Isayev,~O.; Walsh,~A. Machine
  learning for molecular and materials science. \emph{Nature} \textbf{2018},
  \emph{559}, 547\relax
\mciteBstWouldAddEndPuncttrue
\mciteSetBstMidEndSepPunct{\mcitedefaultmidpunct}
{\mcitedefaultendpunct}{\mcitedefaultseppunct}\relax
\EndOfBibitem
\bibitem[Hautier \latin{et~al.}(2010)Hautier, Fischer, Jain, Mueller, and
  Ceder]{ternaryoxcomp}
Hautier,~G.; Fischer,~C.~C.; Jain,~A.; Mueller,~T.; Ceder,~G. Finding
  nature’s missing ternary oxide compounds using machine learning and density
  functional theory. \emph{Chem. Mater.} \textbf{2010}, \emph{22},
  3762--3767\relax
\mciteBstWouldAddEndPuncttrue
\mciteSetBstMidEndSepPunct{\mcitedefaultmidpunct}
{\mcitedefaultendpunct}{\mcitedefaultseppunct}\relax
\EndOfBibitem
\bibitem[Pilania \latin{et~al.}(2013)Pilania, Wang, Jiang, Rajasekaran, and
  Ramprasad]{acceleratingmatdiscov}
Pilania,~G.; Wang,~C.; Jiang,~X.; Rajasekaran,~S.; Ramprasad,~R. Accelerating
  materials property predictions using machine learning. \emph{Sci. Rep.}
  \textbf{2013}, \emph{3}, 2810\relax
\mciteBstWouldAddEndPuncttrue
\mciteSetBstMidEndSepPunct{\mcitedefaultmidpunct}
{\mcitedefaultendpunct}{\mcitedefaultseppunct}\relax
\EndOfBibitem
\bibitem[Meredig \latin{et~al.}(2014)Meredig, Agrawal, Kirklin, Saal, Doak,
  Thompson, Zhang, Choudhary, and Wolverton]{combinatorialscreen}
Meredig,~B.; Agrawal,~A.; Kirklin,~S.; Saal,~J.~E.; Doak,~J.; Thompson,~A.;
  Zhang,~K.; Choudhary,~A.; Wolverton,~C. Combinatorial screening for new
  materials in unconstrained composition space with machine learning.
  \emph{Phys. Rev. B} \textbf{2014}, \emph{89}, 094104\relax
\mciteBstWouldAddEndPuncttrue
\mciteSetBstMidEndSepPunct{\mcitedefaultmidpunct}
{\mcitedefaultendpunct}{\mcitedefaultseppunct}\relax
\EndOfBibitem
\bibitem[Takigawa \latin{et~al.}(2016)Takigawa, Shimizu, Tsuda, and
  Takakusagi]{dbandcen}
Takigawa,~I.; Shimizu,~K.-i.; Tsuda,~K.; Takakusagi,~S. Machine-learning
  prediction of the d-band center for metals and bimetals. \emph{RSC Adv.}
  \textbf{2016}, \emph{6}, 52587--52595\relax
\mciteBstWouldAddEndPuncttrue
\mciteSetBstMidEndSepPunct{\mcitedefaultmidpunct}
{\mcitedefaultendpunct}{\mcitedefaultseppunct}\relax
\EndOfBibitem
\bibitem[Ulissi \latin{et~al.}(2017)Ulissi, Medford, Bligaard, and
  N{\o}rskov]{rxnnetwork}
Ulissi,~Z.~W.; Medford,~A.~J.; Bligaard,~T.; N{\o}rskov,~J.~K. To address
  surface reaction network complexity using scaling relations machine learning
  and DFT calculations. \emph{Nat. Comm.} \textbf{2017}, \emph{8}, 14621\relax
\mciteBstWouldAddEndPuncttrue
\mciteSetBstMidEndSepPunct{\mcitedefaultmidpunct}
{\mcitedefaultendpunct}{\mcitedefaultseppunct}\relax
\EndOfBibitem
\bibitem[Bose \latin{et~al.}(2018)Bose, Dhawan, Nandi, Sarkar, and
  Ghosh]{bose2018machine}
Bose,~S.; Dhawan,~D.; Nandi,~S.; Sarkar,~R.~R.; Ghosh,~D. Machine learning
  prediction of interaction energies in rigid water clusters. \emph{Physical
  Chemistry Chemical Physics} \textbf{2018}, \emph{20}, 22987--22996\relax
\mciteBstWouldAddEndPuncttrue
\mciteSetBstMidEndSepPunct{\mcitedefaultmidpunct}
{\mcitedefaultendpunct}{\mcitedefaultseppunct}\relax
\EndOfBibitem
\bibitem[Bukkapatnam \latin{et~al.}(2006)Bukkapatnam, Malshe, Agrawal, Raff,
  and Komanduri]{bukkapatnam2006parametrization}
Bukkapatnam,~S.; Malshe,~M.; Agrawal,~P.; Raff,~L.; Komanduri,~R.
  Parametrization of interatomic potential functions using a genetic algorithm
  accelerated with a neural network. \emph{Phys. Rev. B} \textbf{2006},
  \emph{74}, 224102\relax
\mciteBstWouldAddEndPuncttrue
\mciteSetBstMidEndSepPunct{\mcitedefaultmidpunct}
{\mcitedefaultendpunct}{\mcitedefaultseppunct}\relax
\EndOfBibitem
\bibitem[Morawietz and Behler(2013)Morawietz, and Behler]{JPCA2013watercluster}
Morawietz,~T.; Behler,~J. A density-functional theory-based neural network
  potential for water clusters including van der waals corrections. \emph{J.
  Phys. Chem. A} \textbf{2013}, \emph{117}, 7356--7366\relax
\mciteBstWouldAddEndPuncttrue
\mciteSetBstMidEndSepPunct{\mcitedefaultmidpunct}
{\mcitedefaultendpunct}{\mcitedefaultseppunct}\relax
\EndOfBibitem
\bibitem[Natarajan \latin{et~al.}(2015)Natarajan, Morawietz, and
  Behler]{RCS2015water}
Natarajan,~S.~K.; Morawietz,~T.; Behler,~J. Representing the potential-energy
  surface of protonated water clusters by high-dimensional neural network
  potentials. \emph{Phys. Chem. Chem. Phys.} \textbf{2015}, \emph{17},
  8356--8371\relax
\mciteBstWouldAddEndPuncttrue
\mciteSetBstMidEndSepPunct{\mcitedefaultmidpunct}
{\mcitedefaultendpunct}{\mcitedefaultseppunct}\relax
\EndOfBibitem
\bibitem[Manzhos \latin{et~al.}(2015)Manzhos, Dawes, and
  Carrington]{QuantChem2015rev}
Manzhos,~S.; Dawes,~R.; Carrington,~T. Neural network-based approaches for
  building high dimensional and quantum dynamics-friendly potential energy
  surfaces. \emph{Int. J. Quantum Chem.} \textbf{2015}, \emph{115},
  1012--1020\relax
\mciteBstWouldAddEndPuncttrue
\mciteSetBstMidEndSepPunct{\mcitedefaultmidpunct}
{\mcitedefaultendpunct}{\mcitedefaultseppunct}\relax
\EndOfBibitem
\bibitem[Kolb \latin{et~al.}(2016)Kolb, Zhao, Li, Jiang, and
  Guo]{JCP2016permutation}
Kolb,~B.; Zhao,~B.; Li,~J.; Jiang,~B.; Guo,~H. Permutation invariant potential
  energy surfaces for polyatomic reactions using atomistic neural networks.
  \emph{J. Chem. Phys.} \textbf{2016}, \emph{144}, 224103\relax
\mciteBstWouldAddEndPuncttrue
\mciteSetBstMidEndSepPunct{\mcitedefaultmidpunct}
{\mcitedefaultendpunct}{\mcitedefaultseppunct}\relax
\EndOfBibitem
\bibitem[Dragoni \latin{et~al.}(2018)Dragoni, Daff, Cs{\'a}nyi, and
  Marzari]{dragoni2018achieving}
Dragoni,~D.; Daff,~T.~D.; Cs{\'a}nyi,~G.; Marzari,~N. Achieving DFT accuracy
  with a machine-learning interatomic potential: Thermomechanics and defects in
  bcc ferromagnetic iron. \emph{Phys. Rev. Mater.} \textbf{2018}, \emph{2},
  013808\relax
\mciteBstWouldAddEndPuncttrue
\mciteSetBstMidEndSepPunct{\mcitedefaultmidpunct}
{\mcitedefaultendpunct}{\mcitedefaultseppunct}\relax
\EndOfBibitem
\bibitem[Jeong \latin{et~al.}(2018)Jeong, Lee, Yoo, Lee, and
  Han]{jeong2018toward}
Jeong,~W.; Lee,~K.; Yoo,~D.; Lee,~D.; Han,~S. Toward Reliable and Transferable
  Machine Learning Potentials: Uniform Training by Overcoming Sampling Bias.
  \emph{J. Phys. Chem. C} \textbf{2018}, \emph{122}, 22790--22795\relax
\mciteBstWouldAddEndPuncttrue
\mciteSetBstMidEndSepPunct{\mcitedefaultmidpunct}
{\mcitedefaultendpunct}{\mcitedefaultseppunct}\relax
\EndOfBibitem
\bibitem[Zhang \latin{et~al.}(2018)Zhang, Khorshidi, Kastlunger, and
  Peterson]{zhang2018potential}
Zhang,~Y.-J.; Khorshidi,~A.; Kastlunger,~G.; Peterson,~A.~A. The potential for
  machine learning in hybrid QM/MM calculations. \emph{J. Chem. Phys.}
  \textbf{2018}, \emph{148}, 241740\relax
\mciteBstWouldAddEndPuncttrue
\mciteSetBstMidEndSepPunct{\mcitedefaultmidpunct}
{\mcitedefaultendpunct}{\mcitedefaultseppunct}\relax
\EndOfBibitem
\bibitem[Le \latin{et~al.}(2012)Le, Epa, Burden, and
  Winkler]{le2012quantitative}
Le,~T.; Epa,~V.~C.; Burden,~F.~R.; Winkler,~D.~A. Quantitative
  structure--property relationship modeling of diverse materials properties.
  \emph{Chem. Rev} \textbf{2012}, \emph{112}, 2889--2919\relax
\mciteBstWouldAddEndPuncttrue
\mciteSetBstMidEndSepPunct{\mcitedefaultmidpunct}
{\mcitedefaultendpunct}{\mcitedefaultseppunct}\relax
\EndOfBibitem
\bibitem[Sch{\"u}tt \latin{et~al.}(2014)Sch{\"u}tt, Glawe, Brockherde, Sanna,
  M{\"u}ller, and Gross]{schutt2014represent}
Sch{\"u}tt,~K.; Glawe,~H.; Brockherde,~F.; Sanna,~A.; M{\"u}ller,~K.; Gross,~E.
  How to represent crystal structures for machine learning: Towards fast
  prediction of electronic properties. \emph{Phys. Rev. B} \textbf{2014},
  \emph{89}, 205118\relax
\mciteBstWouldAddEndPuncttrue
\mciteSetBstMidEndSepPunct{\mcitedefaultmidpunct}
{\mcitedefaultendpunct}{\mcitedefaultseppunct}\relax
\EndOfBibitem
\bibitem[Von~Lilienfeld \latin{et~al.}(2015)Von~Lilienfeld, Ramakrishnan, Rupp,
  and Knoll]{von2015fourier}
Von~Lilienfeld,~O.~A.; Ramakrishnan,~R.; Rupp,~M.; Knoll,~A. Fourier series of
  atomic radial distribution functions: A molecular fingerprint for machine
  learning models of quantum chemical properties. \emph{Int. J. Quantum Chem.}
  \textbf{2015}, \emph{115}, 1084--1093\relax
\mciteBstWouldAddEndPuncttrue
\mciteSetBstMidEndSepPunct{\mcitedefaultmidpunct}
{\mcitedefaultendpunct}{\mcitedefaultseppunct}\relax
\EndOfBibitem
\bibitem[Seko \latin{et~al.}(2017)Seko, Hayashi, Nakayama, Takahashi, and
  Tanaka]{seko2017representation}
Seko,~A.; Hayashi,~H.; Nakayama,~K.; Takahashi,~A.; Tanaka,~I. Representation
  of compounds for machine-learning prediction of physical properties.
  \emph{Phys. Rev. B} \textbf{2017}, \emph{95}, 144110\relax
\mciteBstWouldAddEndPuncttrue
\mciteSetBstMidEndSepPunct{\mcitedefaultmidpunct}
{\mcitedefaultendpunct}{\mcitedefaultseppunct}\relax
\EndOfBibitem
\bibitem[Rupp \latin{et~al.}(2012)Rupp, Tkatchenko, M{\"u}ller, and
  Von~Lilienfeld]{rupp2012fast}
Rupp,~M.; Tkatchenko,~A.; M{\"u}ller,~K.-R.; Von~Lilienfeld,~O.~A. Fast and
  accurate modeling of molecular atomization energies with machine learning.
  \emph{Phys. Rev. Lett.} \textbf{2012}, \emph{108}, 058301\relax
\mciteBstWouldAddEndPuncttrue
\mciteSetBstMidEndSepPunct{\mcitedefaultmidpunct}
{\mcitedefaultendpunct}{\mcitedefaultseppunct}\relax
\EndOfBibitem
\bibitem[Faber \latin{et~al.}(2017)Faber, Hutchison, Huang, Gilmer, Schoenholz,
  Dahl, Vinyals, Kearnes, Riley, and Von~Lilienfeld]{faber2017prediction}
Faber,~F.~A.; Hutchison,~L.; Huang,~B.; Gilmer,~J.; Schoenholz,~S.~S.;
  Dahl,~G.~E.; Vinyals,~O.; Kearnes,~S.; Riley,~P.~F.; Von~Lilienfeld,~O.~A.
  Prediction errors of molecular machine learning models lower than hybrid DFT
  error. \emph{J. Chem. Theory Comput.} \textbf{2017}, \emph{13},
  5255--5264\relax
\mciteBstWouldAddEndPuncttrue
\mciteSetBstMidEndSepPunct{\mcitedefaultmidpunct}
{\mcitedefaultendpunct}{\mcitedefaultseppunct}\relax
\EndOfBibitem
\bibitem[Bart{\'o}k \latin{et~al.}(2017)Bart{\'o}k, De, Poelking, Bernstein,
  Kermode, Cs{\'a}nyi, and Ceriotti]{bartok2017machine}
Bart{\'o}k,~A.~P.; De,~S.; Poelking,~C.; Bernstein,~N.; Kermode,~J.~R.;
  Cs{\'a}nyi,~G.; Ceriotti,~M. Machine learning unifies the modeling of
  materials and molecules. \emph{Sci. Adv} \textbf{2017}, \emph{3},
  e1701816\relax
\mciteBstWouldAddEndPuncttrue
\mciteSetBstMidEndSepPunct{\mcitedefaultmidpunct}
{\mcitedefaultendpunct}{\mcitedefaultseppunct}\relax
\EndOfBibitem
\bibitem[Davran-Candan \latin{et~al.}(2010)Davran-Candan, G{\"u}nay, and
  Y{\i}ld{\i}r{\i}m]{davran2010structure}
Davran-Candan,~T.; G{\"u}nay,~M.~E.; Y{\i}ld{\i}r{\i}m,~R. Structure and
  activity relationship for CO and O 2 adsorption over gold nanoparticles using
  density functional theory and artificial neural networks. \emph{J. Chem.
  Phys.} \textbf{2010}, \emph{132}, 174113\relax
\mciteBstWouldAddEndPuncttrue
\mciteSetBstMidEndSepPunct{\mcitedefaultmidpunct}
{\mcitedefaultendpunct}{\mcitedefaultseppunct}\relax
\EndOfBibitem
\bibitem[J{\"a}ger \latin{et~al.}(2018)J{\"a}ger, Morooka, Canova, Himanen, and
  Foster]{jager2018machine}
J{\"a}ger,~M.~O.; Morooka,~E.~V.; Canova,~F.~F.; Himanen,~L.; Foster,~A.~S.
  Machine learning hydrogen adsorption on nanoclusters through structural
  descriptors. \emph{npj Comput. Mater.} \textbf{2018}, \emph{4}, 37\relax
\mciteBstWouldAddEndPuncttrue
\mciteSetBstMidEndSepPunct{\mcitedefaultmidpunct}
{\mcitedefaultendpunct}{\mcitedefaultseppunct}\relax
\EndOfBibitem
\bibitem[Musil \latin{et~al.}(2018)Musil, De, Yang, Campbell, Day, and
  Ceriotti]{musil2018machine}
Musil,~F.; De,~S.; Yang,~J.; Campbell,~J.~E.; Day,~G.~M.; Ceriotti,~M. Machine
  learning for the structure--energy--property landscapes of molecular
  crystals. \emph{Chem. Sci.} \textbf{2018}, \emph{9}, 1289--1300\relax
\mciteBstWouldAddEndPuncttrue
\mciteSetBstMidEndSepPunct{\mcitedefaultmidpunct}
{\mcitedefaultendpunct}{\mcitedefaultseppunct}\relax
\EndOfBibitem
\bibitem[Hansen \latin{et~al.}(2015)Hansen, Biegler, Ramakrishnan, Pronobis,
  von Lilienfeld, Müller, and Tkatchenko]{hansen2015}
Hansen,~K.; Biegler,~F.; Ramakrishnan,~R.; Pronobis,~W.; von Lilienfeld,~O.~A.;
  Müller,~K.-R.; Tkatchenko,~A. Machine Learning Predictions of Molecular
  Properties: Accurate Many-Body Potentials and Nonlocality in Chemical Space.
  \emph{J. Phys. Chem. Lett.} \textbf{2015}, \emph{6}, 2326--2331\relax
\mciteBstWouldAddEndPuncttrue
\mciteSetBstMidEndSepPunct{\mcitedefaultmidpunct}
{\mcitedefaultendpunct}{\mcitedefaultseppunct}\relax
\EndOfBibitem
\bibitem[Xie and Grossman(2018)Xie, and Grossman]{xie2018crystal}
Xie,~T.; Grossman,~J.~C. Crystal graph convolutional neural networks for an
  accurate and interpretable prediction of material properties. \emph{Phys.
  Rev. Lett.} \textbf{2018}, \emph{120}, 145301\relax
\mciteBstWouldAddEndPuncttrue
\mciteSetBstMidEndSepPunct{\mcitedefaultmidpunct}
{\mcitedefaultendpunct}{\mcitedefaultseppunct}\relax
\EndOfBibitem
\bibitem[Pilania \latin{et~al.}(2013)Pilania, Wang, Jiang, Rajasekaran, and
  Ramprasad]{pilania2013accelerating}
Pilania,~G.; Wang,~C.; Jiang,~X.; Rajasekaran,~S.; Ramprasad,~R. Accelerating
  materials property predictions using machine learning. \emph{Sci. Rep}
  \textbf{2013}, \emph{3}, 2810\relax
\mciteBstWouldAddEndPuncttrue
\mciteSetBstMidEndSepPunct{\mcitedefaultmidpunct}
{\mcitedefaultendpunct}{\mcitedefaultseppunct}\relax
\EndOfBibitem
\bibitem[Montavon \latin{et~al.}(2013)Montavon, Rupp, Gobre,
  Vazquez-Mayagoitia, Hansen, Tkatchenko, M{\"u}ller, and
  Von~Lilienfeld]{montavon2013machine}
Montavon,~G.; Rupp,~M.; Gobre,~V.; Vazquez-Mayagoitia,~A.; Hansen,~K.;
  Tkatchenko,~A.; M{\"u}ller,~K.-R.; Von~Lilienfeld,~O.~A. Machine learning of
  molecular electronic properties in chemical compound space. \emph{New J.
  Phys} \textbf{2013}, \emph{15}, 095003\relax
\mciteBstWouldAddEndPuncttrue
\mciteSetBstMidEndSepPunct{\mcitedefaultmidpunct}
{\mcitedefaultendpunct}{\mcitedefaultseppunct}\relax
\EndOfBibitem
\bibitem[Rupp \latin{et~al.}(2015)Rupp, Ramakrishnan, and von
  Lilienfeld]{rupp2015machine}
Rupp,~M.; Ramakrishnan,~R.; von Lilienfeld,~O.~A. Machine learning for quantum
  mechanical properties of atoms in molecules. \emph{J. Phys. Chem. Lett.}
  \textbf{2015}, \emph{6}, 3309--3313\relax
\mciteBstWouldAddEndPuncttrue
\mciteSetBstMidEndSepPunct{\mcitedefaultmidpunct}
{\mcitedefaultendpunct}{\mcitedefaultseppunct}\relax
\EndOfBibitem
\bibitem[Chandrasekaran \latin{et~al.}(2019)Chandrasekaran, Kamal, Batra, Kim,
  Chen, and Ramprasad]{chandrasekaran2019solving}
Chandrasekaran,~A.; Kamal,~D.; Batra,~R.; Kim,~C.; Chen,~L.; Ramprasad,~R.
  Solving the electronic structure problem with machine learning. \emph{npj
  Comput. Mater.} \textbf{2019}, \emph{5}, 22\relax
\mciteBstWouldAddEndPuncttrue
\mciteSetBstMidEndSepPunct{\mcitedefaultmidpunct}
{\mcitedefaultendpunct}{\mcitedefaultseppunct}\relax
\EndOfBibitem
\bibitem[Hansen \latin{et~al.}(2013)Hansen, Montavon, Biegler, Fazli, Rupp,
  Scheffler, von Lilienfeld, Tkatchenko, and Müller]{hansen2013}
Hansen,~K.; Montavon,~G.; Biegler,~F.; Fazli,~S.; Rupp,~M.; Scheffler,~M.; von
  Lilienfeld,~O.~A.; Tkatchenko,~A.; Müller,~K.-R. Assessment and Validation
  of Machine Learning Methods for Predicting Molecular Atomization Energies.
  \emph{J. Chem. Theory Comput.} \textbf{2013}, \emph{9}, 3404--3419\relax
\mciteBstWouldAddEndPuncttrue
\mciteSetBstMidEndSepPunct{\mcitedefaultmidpunct}
{\mcitedefaultendpunct}{\mcitedefaultseppunct}\relax
\EndOfBibitem
\bibitem[Ma \latin{et~al.}(2015)Ma, Li, Achenie, and Xin]{Co2chemisorp}
Ma,~X.; Li,~Z.; Achenie,~L.~E.; Xin,~H. Machine-learning-augmented
  chemisorption model for CO2 electroreduction catalyst screening. \emph{J.
  Phys. Chem. Lett.} \textbf{2015}, \emph{6}, 3528--3533\relax
\mciteBstWouldAddEndPuncttrue
\mciteSetBstMidEndSepPunct{\mcitedefaultmidpunct}
{\mcitedefaultendpunct}{\mcitedefaultseppunct}\relax
\EndOfBibitem
\bibitem[Li \latin{et~al.}(2017)Li, Wang, Chin, Achenie, and Xin]{bimetcat}
Li,~Z.; Wang,~S.; Chin,~W.~S.; Achenie,~L.~E.; Xin,~H. High-throughput
  screening of bimetallic catalysts enabled by machine learning. \emph{J.
  Mater. Chem. A} \textbf{2017}, \emph{5}, 24131--24138\relax
\mciteBstWouldAddEndPuncttrue
\mciteSetBstMidEndSepPunct{\mcitedefaultmidpunct}
{\mcitedefaultendpunct}{\mcitedefaultseppunct}\relax
\EndOfBibitem
\bibitem[Li \latin{et~al.}(2017)Li, Ma, and Xin]{featureengg}
Li,~Z.; Ma,~X.; Xin,~H. Feature engineering of machine-learning chemisorption
  models for catalyst design. \emph{Catal. Today} \textbf{2017}, \emph{280},
  232--238\relax
\mciteBstWouldAddEndPuncttrue
\mciteSetBstMidEndSepPunct{\mcitedefaultmidpunct}
{\mcitedefaultendpunct}{\mcitedefaultseppunct}\relax
\EndOfBibitem
\bibitem[Andriotis \latin{et~al.}(2014)Andriotis, Mpourmpakis, Broderick,
  Rajan, Datta, Sunkara, and Menon]{informatics}
Andriotis,~A.~N.; Mpourmpakis,~G.; Broderick,~S.; Rajan,~K.; Datta,~S.;
  Sunkara,~M.; Menon,~M. Informatics guided discovery of surface
  structure-chemistry relationships in catalytic nanoparticles. \emph{J. Chem.
  Phys.} \textbf{2014}, \emph{140}, 094705\relax
\mciteBstWouldAddEndPuncttrue
\mciteSetBstMidEndSepPunct{\mcitedefaultmidpunct}
{\mcitedefaultendpunct}{\mcitedefaultseppunct}\relax
\EndOfBibitem
\bibitem[Li \latin{et~al.}(2017)Li, Zhang, and Liu]{li2017application}
Li,~H.; Zhang,~Z.; Liu,~Z. Application of artificial neural networks for
  catalysis: A review. \emph{Catalysts} \textbf{2017}, \emph{7}, 306\relax
\mciteBstWouldAddEndPuncttrue
\mciteSetBstMidEndSepPunct{\mcitedefaultmidpunct}
{\mcitedefaultendpunct}{\mcitedefaultseppunct}\relax
\EndOfBibitem
\bibitem[Jinnouchi and Asahi(2017)Jinnouchi, and Asahi]{predcatnano}
Jinnouchi,~R.; Asahi,~R. Predicting Catalytic Activity of Nanoparticles by a
  DFT-Aided Machine-Learning Algorithm. \emph{J. Phys. Chem. Lett.}
  \textbf{2017}, \emph{8}, 4279--4283\relax
\mciteBstWouldAddEndPuncttrue
\mciteSetBstMidEndSepPunct{\mcitedefaultmidpunct}
{\mcitedefaultendpunct}{\mcitedefaultseppunct}\relax
\EndOfBibitem
\bibitem[Gasper \latin{et~al.}(2017)Gasper, Shi, and Ramasubramaniam]{CoonPt}
Gasper,~R.; Shi,~H.; Ramasubramaniam,~A. Adsorption of co on low-energy,
  low-symmetry pt nanoparticles: Energy decomposition analysis and prediction
  via machine-learning models. \emph{J. Phys. Chem. C} \textbf{2017},
  \emph{121}, 5612--5619\relax
\mciteBstWouldAddEndPuncttrue
\mciteSetBstMidEndSepPunct{\mcitedefaultmidpunct}
{\mcitedefaultendpunct}{\mcitedefaultseppunct}\relax
\EndOfBibitem
\bibitem[Kitchin(2018)]{kitchin2018machine}
Kitchin,~J.~R. Machine learning in catalysis. \emph{Nat. Catal.} \textbf{2018},
  \emph{1}, 230\relax
\mciteBstWouldAddEndPuncttrue
\mciteSetBstMidEndSepPunct{\mcitedefaultmidpunct}
{\mcitedefaultendpunct}{\mcitedefaultseppunct}\relax
\EndOfBibitem
\bibitem[Neal \latin{et~al.}(2007)Neal, Starace, Jarrold, Joshi, Krishnamurty,
  and Kanhere]{neal2007melting}
Neal,~C.~M.; Starace,~A.~K.; Jarrold,~M.~F.; Joshi,~K.; Krishnamurty,~S.;
  Kanhere,~D.~G. Melting of Aluminum Cluster Cations with 31- 48 Atoms:
  Experiment and Theory. \emph{The Journal of Physical Chemistry C}
  \textbf{2007}, \emph{111}, 17788--17794\relax
\mciteBstWouldAddEndPuncttrue
\mciteSetBstMidEndSepPunct{\mcitedefaultmidpunct}
{\mcitedefaultendpunct}{\mcitedefaultseppunct}\relax
\EndOfBibitem
\bibitem[Starace \latin{et~al.}(2008)Starace, Neal, Cao, Jarrold, Aguado, and
  L{\'o}pez]{aguado2008}
Starace,~A.~K.; Neal,~C.~M.; Cao,~B.; Jarrold,~M.~F.; Aguado,~A.;
  L{\'o}pez,~J.~M. Correlation between the latent heats and cohesive energies
  of metal clusters. \emph{J. Chem. Phys.} \textbf{2008}, \emph{129},
  144702\relax
\mciteBstWouldAddEndPuncttrue
\mciteSetBstMidEndSepPunct{\mcitedefaultmidpunct}
{\mcitedefaultendpunct}{\mcitedefaultseppunct}\relax
\EndOfBibitem
\bibitem[Aguado and L{\'o}pez(2009)Aguado, and L{\'o}pez]{aguado2009}
Aguado,~A.; L{\'o}pez,~J.~M. Structures and stabilities of Al n+, Al n, and Al
  n-(n= 13--34) clusters. \emph{J. Chem. Phys.} \textbf{2009}, \emph{130},
  064704\relax
\mciteBstWouldAddEndPuncttrue
\mciteSetBstMidEndSepPunct{\mcitedefaultmidpunct}
{\mcitedefaultendpunct}{\mcitedefaultseppunct}\relax
\EndOfBibitem
\bibitem[Susan and Joshi(2014)Susan, and Joshi]{anju2014}
Susan,~A.; Joshi,~K. Rationalizing the role of structural motif and underlying
  electronic structure in the finite temperature behavior of atomic clusters.
  \emph{J. Chem. Phys.} \textbf{2014}, \emph{140}, 154307\relax
\mciteBstWouldAddEndPuncttrue
\mciteSetBstMidEndSepPunct{\mcitedefaultmidpunct}
{\mcitedefaultendpunct}{\mcitedefaultseppunct}\relax
\EndOfBibitem
\bibitem[Bl\"ochl(1994)]{paw1}
Bl\"ochl,~P.~E. Projector augmented-wave method. \emph{Phys. Rev. B}
  \textbf{1994}, \emph{50}, 17953--17979\relax
\mciteBstWouldAddEndPuncttrue
\mciteSetBstMidEndSepPunct{\mcitedefaultmidpunct}
{\mcitedefaultendpunct}{\mcitedefaultseppunct}\relax
\EndOfBibitem
\bibitem[Kresse and Joubert(1999)Kresse, and Joubert]{paw2}
Kresse,~G.; Joubert,~D. From ultrasoft pseudopotentials to the projector
  augmented-wave method. \emph{Phys. Rev. B} \textbf{1999}, \emph{59},
  1758--1775\relax
\mciteBstWouldAddEndPuncttrue
\mciteSetBstMidEndSepPunct{\mcitedefaultmidpunct}
{\mcitedefaultendpunct}{\mcitedefaultseppunct}\relax
\EndOfBibitem
\bibitem[Perdew \latin{et~al.}(1996)Perdew, Burke, and Ernzerhof]{pbe1}
Perdew,~J.~P.; Burke,~K.; Ernzerhof,~M. Generalized Gradient Approximation Made
  Simple. \emph{Phys. Rev. Lett.} \textbf{1996}, \emph{77}, 3865--3868\relax
\mciteBstWouldAddEndPuncttrue
\mciteSetBstMidEndSepPunct{\mcitedefaultmidpunct}
{\mcitedefaultendpunct}{\mcitedefaultseppunct}\relax
\EndOfBibitem
\bibitem[Perdew \latin{et~al.}(1997)Perdew, Burke, and Ernzerhof]{pbe2}
Perdew,~J.~P.; Burke,~K.; Ernzerhof,~M. Generalized Gradient Approximation Made
  Simple [Phys. Rev. Lett. 77, 3865 (1996)]. \emph{Phys. Rev. Lett.}
  \textbf{1997}, \emph{78}, 1396--1396\relax
\mciteBstWouldAddEndPuncttrue
\mciteSetBstMidEndSepPunct{\mcitedefaultmidpunct}
{\mcitedefaultendpunct}{\mcitedefaultseppunct}\relax
\EndOfBibitem
\bibitem[Kresse and Hafner(1994)Kresse, and Hafner]{vasp1}
Kresse,~G.; Hafner,~J. \textit{Ab initio} molecular-dynamics simulation of the
  liquid-metal--amorphous-semiconductor transition in germanium. \emph{Phys.
  Rev. B} \textbf{1994}, \emph{49}, 14251--14269\relax
\mciteBstWouldAddEndPuncttrue
\mciteSetBstMidEndSepPunct{\mcitedefaultmidpunct}
{\mcitedefaultendpunct}{\mcitedefaultseppunct}\relax
\EndOfBibitem
\bibitem[Kresse and Furthm\"uller(1996)Kresse, and Furthm\"uller]{vasp2}
Kresse,~G.; Furthm\"uller,~J. Efficient iterative schemes for \textit{ab
  initio} total-energy calculations using a plane-wave basis set. \emph{Phys.
  Rev. B} \textbf{1996}, \emph{54}, 11169--11186\relax
\mciteBstWouldAddEndPuncttrue
\mciteSetBstMidEndSepPunct{\mcitedefaultmidpunct}
{\mcitedefaultendpunct}{\mcitedefaultseppunct}\relax
\EndOfBibitem
\bibitem[Kresse and Furthmüller(1996)Kresse, and Furthmüller]{vasp3}
Kresse,~G.; Furthmüller,~J. Efficiency of ab-initio total energy calculations
  for metals and semiconductors using a plane-wave basis set. \emph{Comput.
  Mater. Sci.} \textbf{1996}, \emph{6}, 15 -- 50\relax
\mciteBstWouldAddEndPuncttrue
\mciteSetBstMidEndSepPunct{\mcitedefaultmidpunct}
{\mcitedefaultendpunct}{\mcitedefaultseppunct}\relax
\EndOfBibitem
\bibitem[Pedregosa \latin{et~al.}(2011)Pedregosa, Varoquaux, Gramfort, Michel,
  Thirion, Grisel, Blondel, Prettenhofer, Weiss, Dubourg, Vanderplas, Passos,
  Cournapeau, Brucher, Perrot, and Duchesnay]{scikit}
Pedregosa,~F.; Varoquaux,~G.; Gramfort,~A.; Michel,~V.; Thirion,~B.;
  Grisel,~O.; Blondel,~M.; Prettenhofer,~P.; Weiss,~R.; Dubourg,~V.
  \latin{et~al.}  Scikit-learn: Machine Learning in {P}ython. \emph{J. Mach.
  Learn. Res.} \textbf{2011}, \emph{12}, 2825--2830\relax
\mciteBstWouldAddEndPuncttrue
\mciteSetBstMidEndSepPunct{\mcitedefaultmidpunct}
{\mcitedefaultendpunct}{\mcitedefaultseppunct}\relax
\EndOfBibitem
\end{mcitethebibliography}
\bibliographystyle{acs}
\end{document}